%% file: main.tex
\documentclass[a4paper,11pt]{article}
\pdfoutput=1 

\usepackage{jcappub} 

\usepackage[T1]{fontenc} 
\usepackage[utf8]{inputenc}
\usepackage{aas_macros}
\usepackage{comment}
\usepackage[colorinlistoftodos]{todonotes}
\usepackage{tikz}
\usepackage{float}
\usetikzlibrary{arrows,shadows,petri,decorations.markings,shapes}
\usepackage{standalone}

\title{\boldmath Dark Matter Subhalos, Strong Lensing and Machine Learning}

\author[]{Sreedevi Varma,}
\author[]{Malcolm Fairbairn \&}
\author[]{Julio Figueroa}

\affiliation[]{Theoretical Particle Physics and Cosmology, \\Physics, King's College London,\\ London WC2R 2LS, \\ United Kingdom}

\emailAdd{sreedevi.varma@kcl.ac.uk}
\emailAdd{malcolm.fairbairn@kcl.ac.uk}
\emailAdd{julio.figueroa@kcl.ac.uk}

\abstract{We investigate the possibility of applying machine learning techniques to images of strongly lensed galaxies to detect a low  mass cut-off in the spectrum of dark matter sub-halos within the lens system.  We generate lensed images of systems containing substructure in seven different categories corresponding to lower mass cut-offs ranging from $10^{9}M_\odot$ down to $10^{6}M_\odot$.  We use convolutional neural networks to perform a multi-classification sorting of these images and see that the algorithm is able to correctly identify the lower mass cut-off within an order of magnitude to better than 93\% accuracy.
\\
\\
\\
\\
KCL-PH-TH/2020-26}

\begin{document}
\maketitle
\flushbottom

\section{Introduction}

The quest to uncover the fundamental nature of Dark Matter (DM)  is an ongoing struggle at the intersection of Astrophysics, Particle Physics and Cosmology. No direct evidence for the existence of DM has been obtained so far from terrestrial \cite{Akerib:2016vxi,Cui:2016ppb,Aprile:2018dbl} astrophysical \cite{Aprile:2018dbl,Chang:2018bpt,Lisanti:2017qlb,Fermi-LAT:2016uux} or collider searches \cite{Sirunyan:2016iap,Aaboud:2019yqu} which makes gravitational effects deduced from astrophysical observations the only avenue to probe the properties of this invisible particle \cite{Brehmer:2019jyt}. 

The widely accepted and well defined paradigm of Cold Dark Matter in the context of the $\Lambda$\textsc{CDM} model describes DM as a cold slow-moving and collision-less particle which emits no electromagnetic radiation. Observations of the universe varying from galaxy clustering \cite{Alam:2016hwk}, supernova luminosities \cite{Scolnic:2017caz} and the shape of Cosmic Microwave Background (CMB) correlation function \cite{Aghanim:2018eyx} are all well explained by the $\Lambda$\textsc{CDM} framework \cite{DiazRivero:2019hxf}. The model provides an excellent description of the observed distribution of the matter on large scales and predicts a hierarchical growth of structures (small structures collapse under their own self gravity and subsequently merge to form consecutively larger structures) - a paradigm which is supported by various observations \cite{Aghanim:2018eyx,10.1093/mnras/stz2310,Petac:2019lam}.  This means that large dark matter halos like galaxies should be composed of many smaller virialised halos with progressively smaller masses, spanning many orders of magnitude in mass \cite{Diemand:2005vz}.

 Since we can only measure dark matter through its gravitational effect upon baryons, most probes of its properties only exist on mass scales equal to or above that of the regions which collapse to form the smallest galaxies containing stars.  Alternatives to CDM like warm DM \cite{PhysRevLett.72.17,Bode_2001, PhysRevD.76.103515}, self-interacting DM \cite{Tulin:2017ara} and fuzzy DM \cite{PhysRevLett.85.1158,PhysRevD.95.043541} behave differently on small scales \cite{Aghanim:2018eyx}.  In particular they predict a lack of smaller galaxies since dark matter halos would be unable to form below characteristic scales defined by the underlying particle physics of the dark matter.  This would correspond to the free streaming scale in the case of warm dark matter \cite{2001ApJ...556...93B} and similar effects on the halo mass fuction occur for ultralight or fuzzy dark matter \cite{Du:2016zcv,Safarzadeh:2019sre}.  The effect of self interacting dark matter upon the number of smaller mass sub halos is less clear \cite{Rocha:2012jg}.  It is however of paramount importance to identify the presence or absence of smaller mass dark matter halos.

The dark matter dominated low mass end of the halo mass function is challenging to probe due a lack of stars, in fact the smallest halo we expect to contain Baryons is around $10^9 M_\odot$ \cite{Sawala:2014hqa}. Nevertheless, there have been suggestions that there are anomalies on small scales which might hint at behaviour beyond cold dark matter paradigm \cite{Bullock:2017xww}.  For example, it has been claimed that dwarf galaxies may possess cores \cite{Walker:2011zu}, but this depends upon Jeans modelling \cite{Richardson:2014mra} and may also be related to observational perspective effects \cite{2018MNRAS.474.1398G} and the history of star formation in dwarf galaxies \cite{Read:2018fxs, Genina:2019job}.  There have  also been claims that there are a lack of satellite galaxies observed in the local group relative to the number predicted in N-body simulations of dark matter \cite{BoylanKolchin:2011dk}, although recently observations have shown the presence of many more of these small galaxies \cite{Drlica-Wagner:2015ufc} while re-analysis of the problem has found different conclusions \cite{Read:2018gpi}. 

It is therefore imperative to find ways to measure the presence or absence of very small (<$10^8 M_\odot$) dark matter halos in galaxies.  One promising avenue which has been pursued in some detail is that of streams of stars being perturbed by smaller subhalos tidally disrupting their structure \cite{Bovy:2016irg} and there are tentative indications of halos as small in mass as $10^7M_\odot$ being detected in perturbed streams around the Milky Way \cite{Banik:2019cza}.  In this work, we will be looking at the possibility of detecting substructure in the dark matter of galactic halos by the effect of such subhalos upon strong lensing signals.  We are not the first people to consider this so let us first look at previous work in this direction.

\subsection{Previous Related Work by other Authors}

Gravitational lensing, a direct consequence of General Relativity, is the deflection of light rays in the presence of massive bodies. It has emerged as a powerful tool in Cosmology and Astrophysics since its observational confirmation on galactic scales in 1979 \cite{1979Natur.279..381W}. Lensing can be used to measure the distribution of mass at a higher redshifts spanning a wide range of scales in the Universe.  Multiple images of a source are produced by what is known as a strong lens system, weak lenses only produce small distortions in the shape of faraway sources and microlensing can amplify the light curves of the source.

The gravitational lensing methods detailed in \cite{Koopmans:2005nr,Vegetti:2008eg} are used to attempt to detect dark substructure by studying small perturbations on the highly magnified Einstein rings and arcs produced by strong lensing by galaxies.
These techniques were shown to be able to detect mass substructure down to $10^8 M_{\odot}$. The authors of \cite{Mao:1997ek,2001ApJ...563....9M} proposed that the flux ratios in the lens system are affected by the substructure in the lens galaxy by analysing the gravitationally lensed images of a quasar lying behind it.  Other substructure detection methods using gravitationally lensed quasars include \cite{Hsueh:2019ynk,Moustakas:2002iz}. In all of these analyses, it is neccesary to model the smooth component of the lens galaxy before doing inference on the substructure properties \cite{DiazRivero:2019hxf}.

Statistical methods which have been used to explore the substructure in lensed images included measuring the power spectrum and applying Bayesian analysis to the lensed images/systems. \cite{Hezaveh:2014aoa} measures the correlation between shape and amplitude of the power spectrum with the abundance, density profile and mass function of subhalos using lensed image residuals. Two point correlation functions are used in \cite{Rivero:2017mao} to study substructure convergence power spectrum for CDM and self interacting dark matter (SIDM) populations. References \cite{Hezaveh:2014aoa,Rivero:2017mao} also study how the subhalo abundance, mass function,  internal density profile, and concentration affect the amplitude and shape of the substructure power spectrum. The substructure power spectrum and how it is related to redshift is explored using high-resolution simulations in \cite{Rivero:2018bcd}. The authors of \cite{Brennan:2018jhq} measure the lensing convergence power spectrum in simulated galaxies using semi-analytic subhalo populations. They propose that the slope and amplitude of the power spectrum are good parameters to distinguish between various dark matter models. 

Bayesian inference can also be used to quantify the substructure content and to classify between different dark matter models. \cite{Birrer:2017rpp} use Approximate Bayesian Computing on the quadrupole lens system of a quasar (RXJ1131-1231) and report a lower limit for the mass of thermal relic dark matter of $m_{TH} =
2keV$ at 2$\sigma$ confidence level. The density profile, number of subhalos and their properties are explored using a Bayesian approach in \cite{Brewer:2015yya}. Other pieces of literature which use statistical methods in computing the distribution of lensing potentials, modelling of strongly lensed systems  and substructure detection using likelihood ratio estimators are \cite{PhysRevD.94.043505}, \cite{Daylan_2018} and \cite{Brehmer:2019jyt} respectively.
The statistical approach in general studies the residuals obtained between the data and smooth lens models in order to infer parameters of the subhalo population \cite{DiazRivero:2019hxf}.

\subsection{Machine Learning approaches}

In recent years, there has been an increasing interest in applying Machine Learning to physics ranging from quantum physics to cosmology \cite{Carleo:2019ptp, Mehta_2019}. Applications of machine learning to the identification of substructure probing include  \cite{Brehmer:2019jyt,DiazRivero:2019hxf,Petac:2019lam,Chianese:2019ifk,Alexander:2019puy}. Reference \cite{Brehmer:2019jyt} trains a neural network to estimate the likelihood function for the detection of subhalos whereas \cite{DiazRivero:2019hxf, Alexander:2019puy,Petac:2019lam} use image recognition algorithms such as Convolutional Neural Network (CNN) to study the dark subhalos. The authors of \cite{Bernardini:2019bmd} propose an approach to predict dark matter halo formation using Deep learning.

Reference \cite{Chianese:2019ifk} presents a fully differentiable lensing pipeline which combines variational autoencoder (VAE)-based source and physical lens which could be useful in probing subhalo effects computationally cheaply.

Lensed images are used to infer likelihood ratio estimates of substructure parameters (like mass fraction $f_{sub}$ and slope  $\beta$ of the mass function) in \cite{Brehmer:2019jyt}. 

The work in \cite{Petac:2019lam} uses a 3D CNN to find the subhalo-induced patterns in spatial and velocity distributions of stars. The network gained an accuracy of $97.1\%$ in classifying uniformly distributed subhalos with the mass in a range of $[3\times10^6,\ 10^7]\ M_{\odot}$. The CNN fails to train the maps below this $3\times\ 10^6\ M_{\odot}$. 

Various CNN architectures are used in \cite{Alexander:2019puy} to classify different substructure classes from lensed images to find the position, mass and other properties of substructures. The multi-class classifier classifies the lensed images into `vortex', `spherical' and `no- substructure' images with a fixed number of subhalos (25). The images are classified into the classes with an area-under-curve (AUC) scores of 0.998, 0.985 and 0.968 for the three classes respectively. 

In \cite{DiazRivero:2019hxf}, the authors propose three methods in using CNN for subhalo detection in the vicinity of an Einstein ring. The lensed images are generated with a single subhalo, with subhalo sample drawn from a Gaussian distribution and fixing the mass fraction with varying the number of subhalos. All these images are classified with images of the host lens without any substructure. Their network was able to probe subhalos of mass $\geq 5\times 10^9 M_{\odot}$ with an accuracy of $75 \%$. Other recent machine learning applications in the strong lensing regime include \cite{Cheng:2019hpp, Shirasaki:2019wxk, Madireddy:2019nrh, 2018A&A...611A...2S, 2019MNRAS.487.5263D, 2017Natur.548..555H, 2017ApJ...850L...7P, 2018arXiv180800011M}.

\subsection{Our Approach}

In our work, we have a different approach to other efforts in the literature.  First, rather than attempting to look at one particular single mass of subhalos, we only consider distributions of subhalos of different masses which represent accurately the distribution observed in N-body simulations.

We employ widely used image recognition algorithms to classifying between lensing images generated for different mass ranges of subhalos. The mass ranges we considered were $10^6 M_{\odot}- 10^{13} M_{\odot},\ 10^{6.5} M_{\odot}- 10^{13} M_{\odot},\ 10^7 M_{\odot}- 10^{13} M_{\odot},\ 10^{7.5} M_{\odot}- 10^{13} M_{\odot},\ 10^8 M_{\odot}- 10^{13} M_{\odot},\ 10^{8.5} M_{\odot}- 10^{13} M_{\odot}$ and $10^9 M_{\odot}- 10^{13} M_{\odot}$ so that all of the distributions have different lower mass cut-offs. The subhalos are populated across the host halos according to the best available predictions for their spatial distribution.  We then use multilabel-classification to determine between the mass ranges.

The remaining part of the paper is organised as follows: Section \ref{sec:data} describes the generation of the lensing data sets which we will subsequently attempt to distinguish between.  Section \ref{sec:ml} will explain the different CNN architectures used in the study. In Section \ref{sec:results} we discuss the results, conclusions and outlook for the future works.
\label{sec:intro}

\section{Data Generation}
\label{sec:data}
We would like to create simulated images of strongly lensed systems with different amounts of sub structure and then see if we are able to identify different situations from each other using machine learning algorithms.  In order to do this, we look at seven different situations, in particular when the subhalo lower mass cut-off corresponds to $10^9$, $10^{8.5}$, $10^8$, $10^{7.5}$, $10^7$, $10^{6.5}$ and $10^6M_\odot$ respectively.  We also want to vary the properties of the source as well as the lens, before generating the lensed images.

\subsection{Properties of the Strong Lens Models\label{subsec:lens}}
We assume that the total mass of the lensing system (mass of host halo + mass of the subhalos) is kept to a constant ($1.10 \times 10^{13} M_{\odot}$).  In order to create a strong lens, given that we are not including baryons in the lens, we have to give the main halo quite a high concentration parameter of $c=15$. 
The subhalos are generated with minimum mass varying from $10^{6} M_{\odot}$ to $10^{9} M_{\odot}$ . The masses of the subhalos are simulated from the halo mass function, as given by the expression 
\begin{equation}
\frac{dn(z)}{d\ln m}\propto\left(\frac{m}{M_\odot}\right)^{A} \exp\left[-B \left(\frac{m}{M}\right)^3\right]
\label{eq:pdf}
\end{equation}
where we use the values from \cite{2010MNRAS.404..502G,Despali:2016meh}. Here, $M$ is the host halo mass and $m$ is the subhalo mass. The parameters $A$ (-0.9) and $B$ (12.2715) represent the slope and exponential cutoff parameter respectively.  We generate halos which form a significant fraction of the mass of the halo, the fraction is consistent with the results from \cite{2010MNRAS.404..502G} and we put the number of halos and the fractional mass in subhalos in table \ref{tablehalo}.  Because we randomly sample the distribution (\ref{eq:pdf}) we have a range of number of subhalos for each mass range.
\begin{table}
\begin{center}
\begin{tabular}{ |c|c|c| } 
 \hline
Mass range & Average number of subhalos & Fractional mass in subhalos  \\ 
 \hline
\ 
$10^6 M_{\odot}-\ 10^{13} M_{\odot}$ & $4.11\times 10^4$ & 0.1 \\ 
$10^{6.5} M_{\odot}-\ 10^{13} M_{\odot}$ & $1.39\times 10^4$ & 0.0965 \\ 
$10^7 M_{\odot}-\ 10^{13} M_{\odot}$ & $5.00\times 10^3$ & 0.0927 \\
$10^{7.5} M_{\odot}-\ 10^{13} M_{\odot}$ & $1.80\times 10^3$ & 0.0875 \\ 

$10^8 M_{\odot}-\ 10^{13} M_{\odot}$ & 659  & 0.0834\\
$10^{8.5} M_{\odot}-\ 10^{13} M_{\odot}$ & 244 & 0.0784  \\ 

$10^9 M_{\odot}-\ 10^{13} M_{\odot}$ & 83.4 & 0.0731 \\
 \hline
 \end{tabular}
 
 \caption{Average number of subhalos for each mass range and their fraction of the mass of the overall halo.\label{tablehalo}}
\end{center}

\end{table}

\begin{figure}[!tbp]
  \centering
  \begin{minipage}[b]{0.45\textwidth}
    \includegraphics[width=\textwidth]{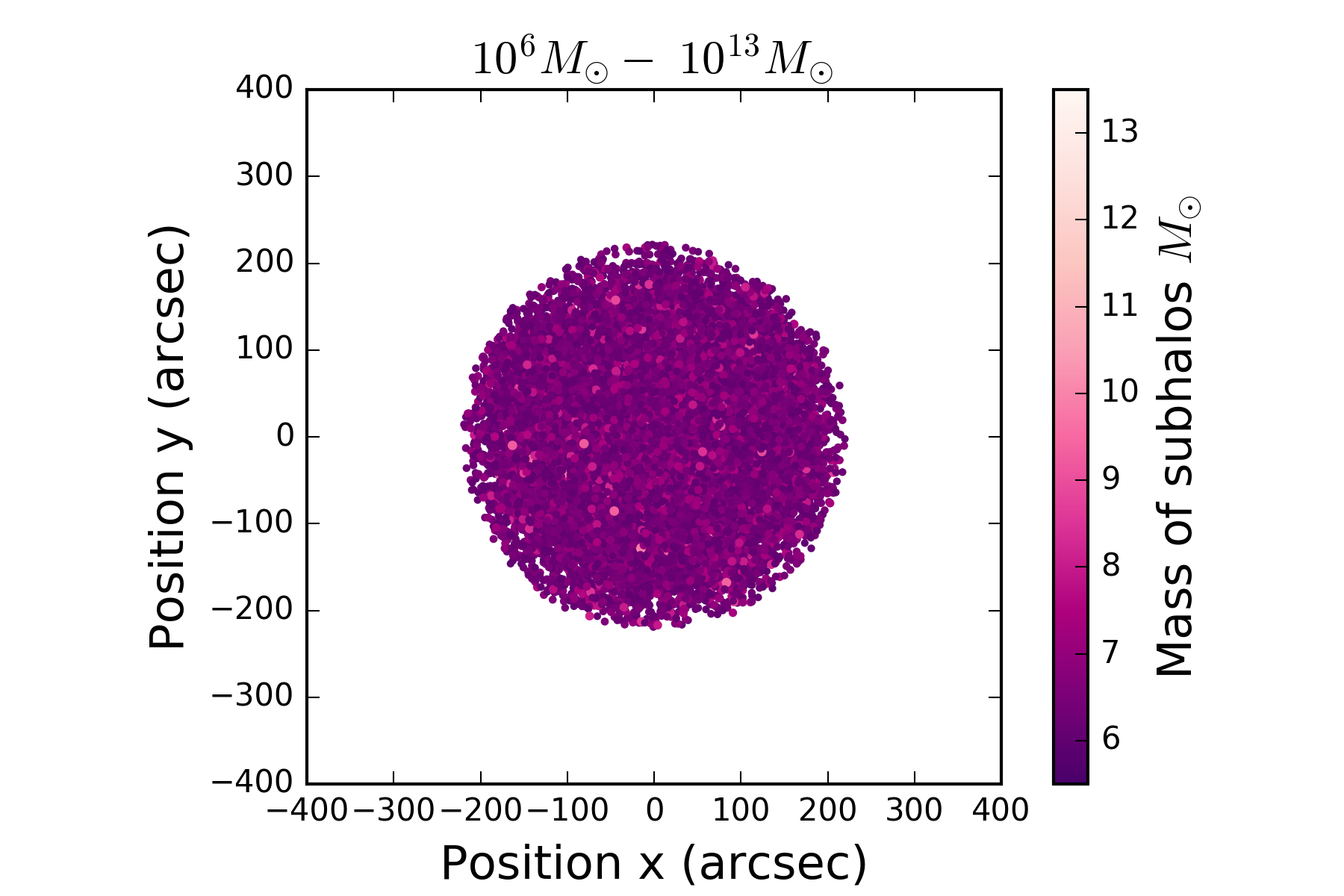}
  
  \end{minipage}
  \hfill
   \begin{minipage}[b]{0.45\textwidth}
    \includegraphics[width=\textwidth]{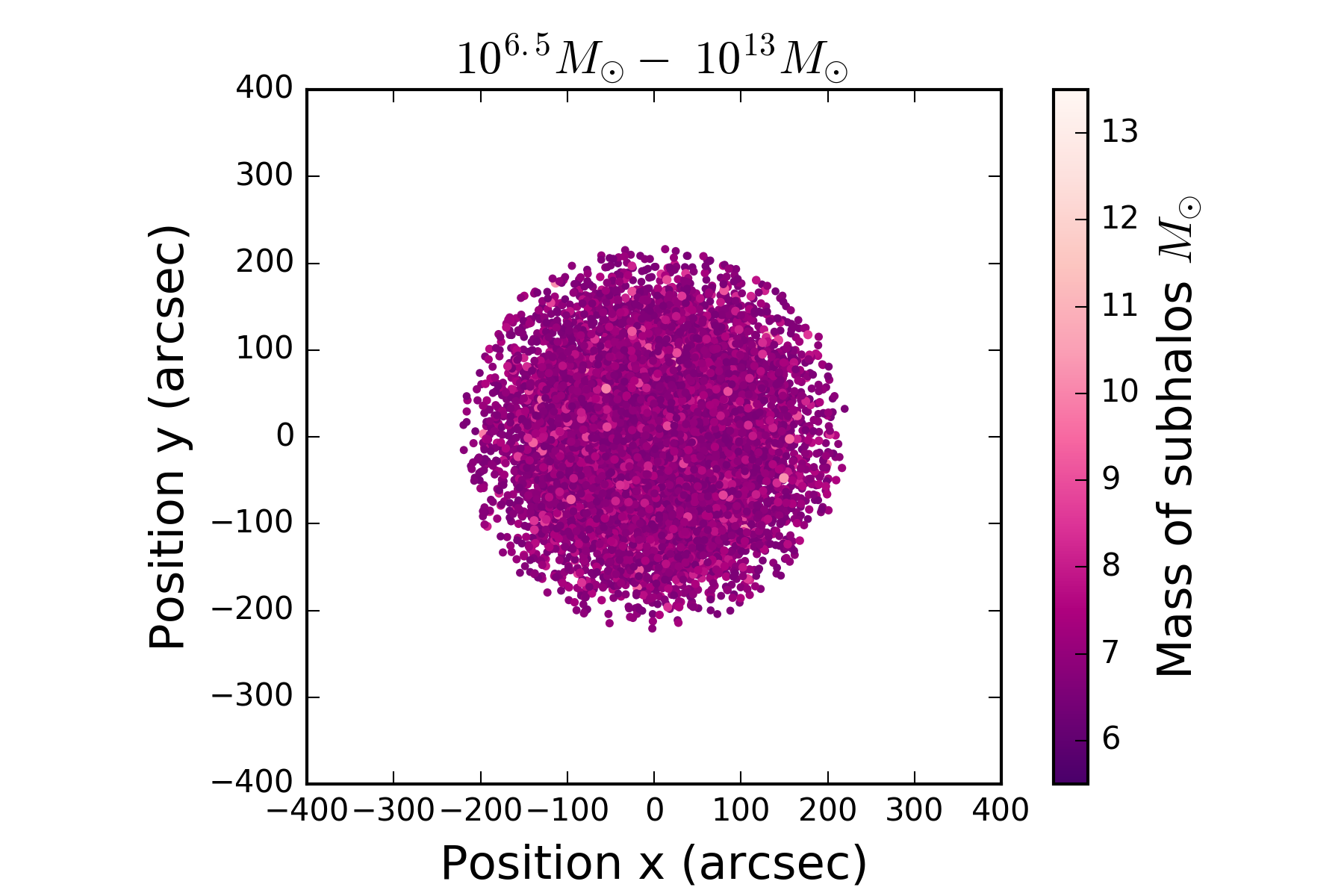}
  
  \end{minipage}
  \hfill
  \begin{minipage}[b]{0.45\textwidth}
    \includegraphics[width=\textwidth]{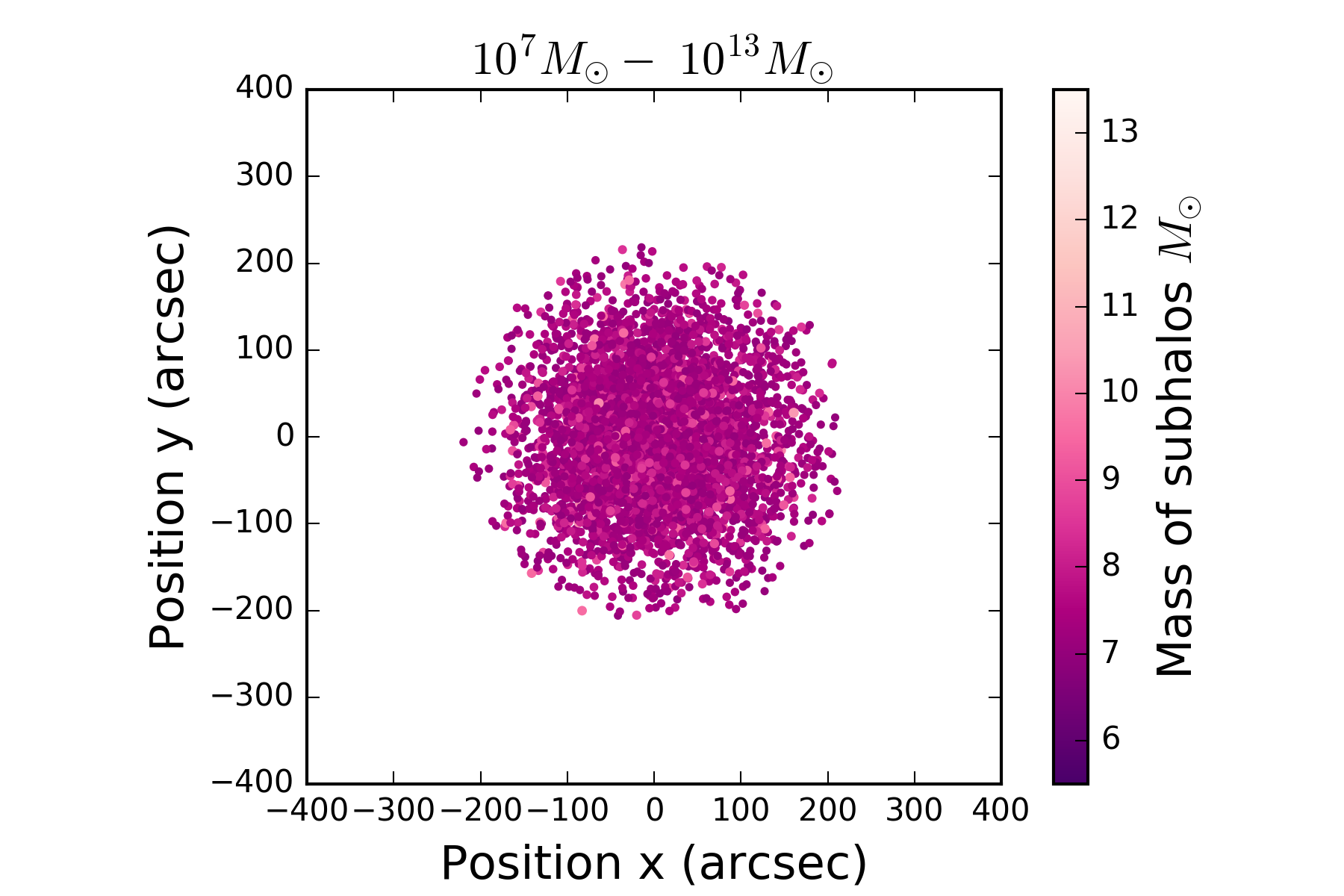}
   
  \end{minipage}
    \hfill
 \begin{minipage}[b]{0.45\textwidth}
    \includegraphics[width=\textwidth]{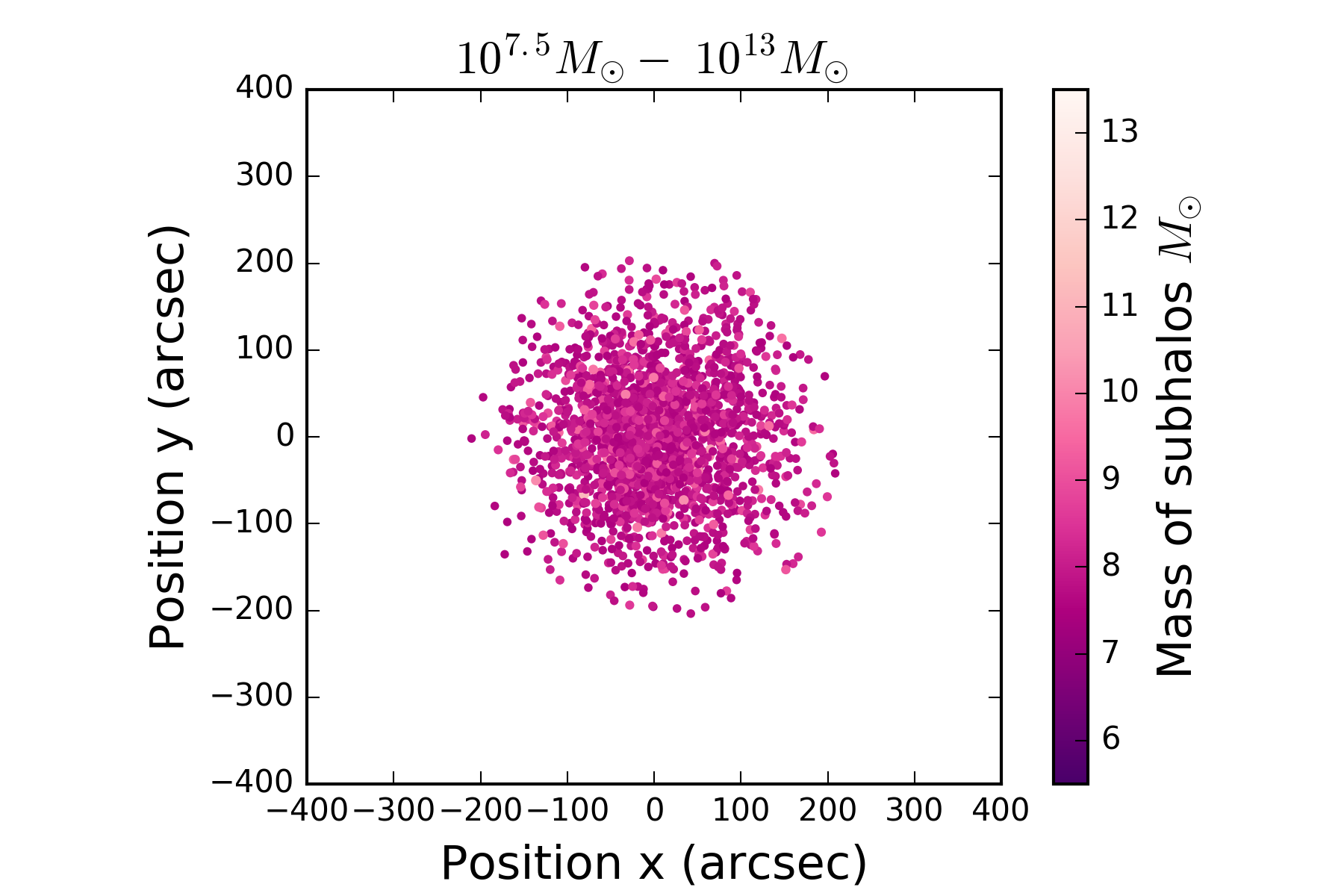}
  
  \end{minipage}
  \hfill
  \begin{minipage}[b]{0.45\textwidth}
    \includegraphics[width=\textwidth]{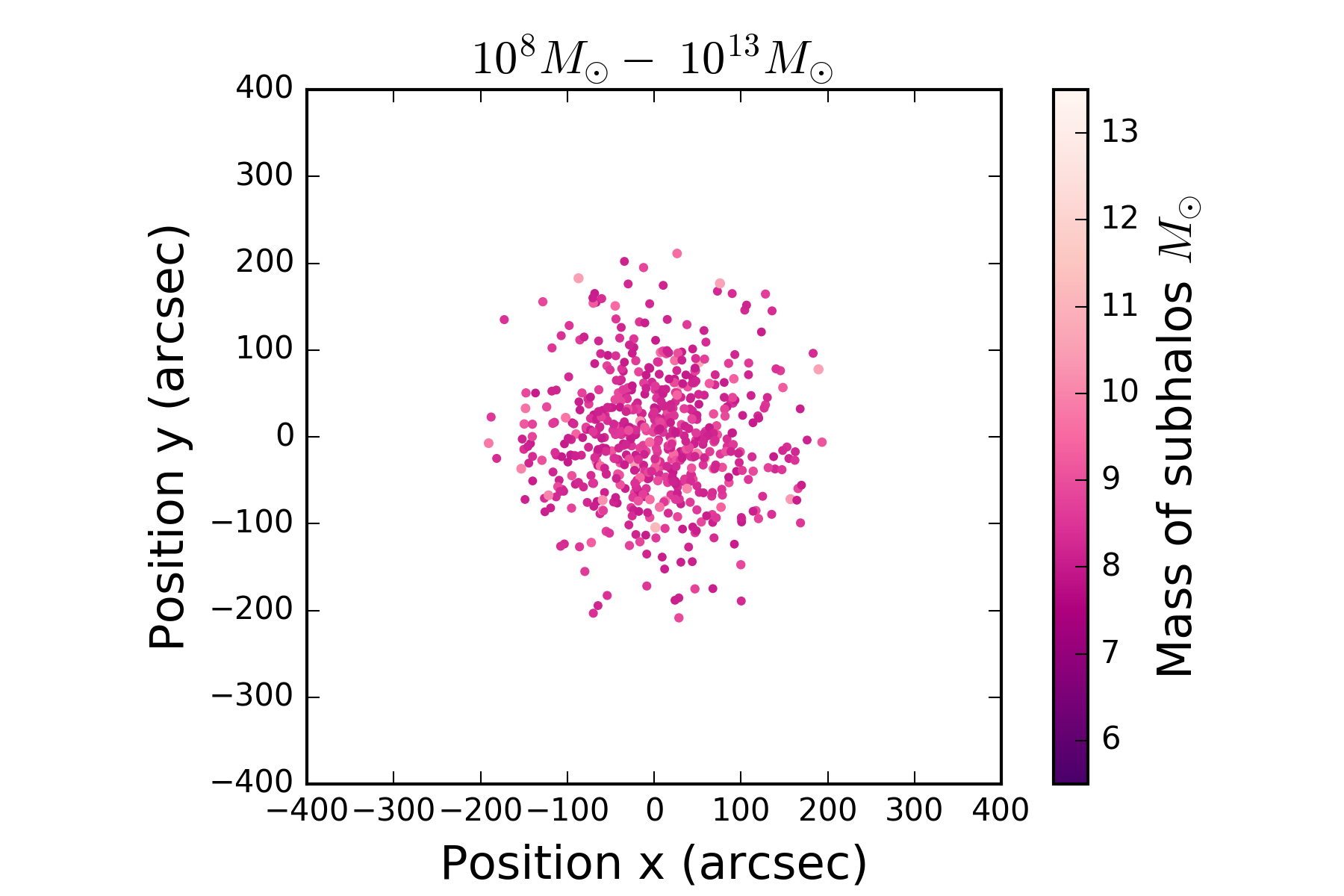}
     \end{minipage}
  \hfill
   \begin{minipage}[b]{0.45\textwidth}
    \includegraphics[width=\textwidth]{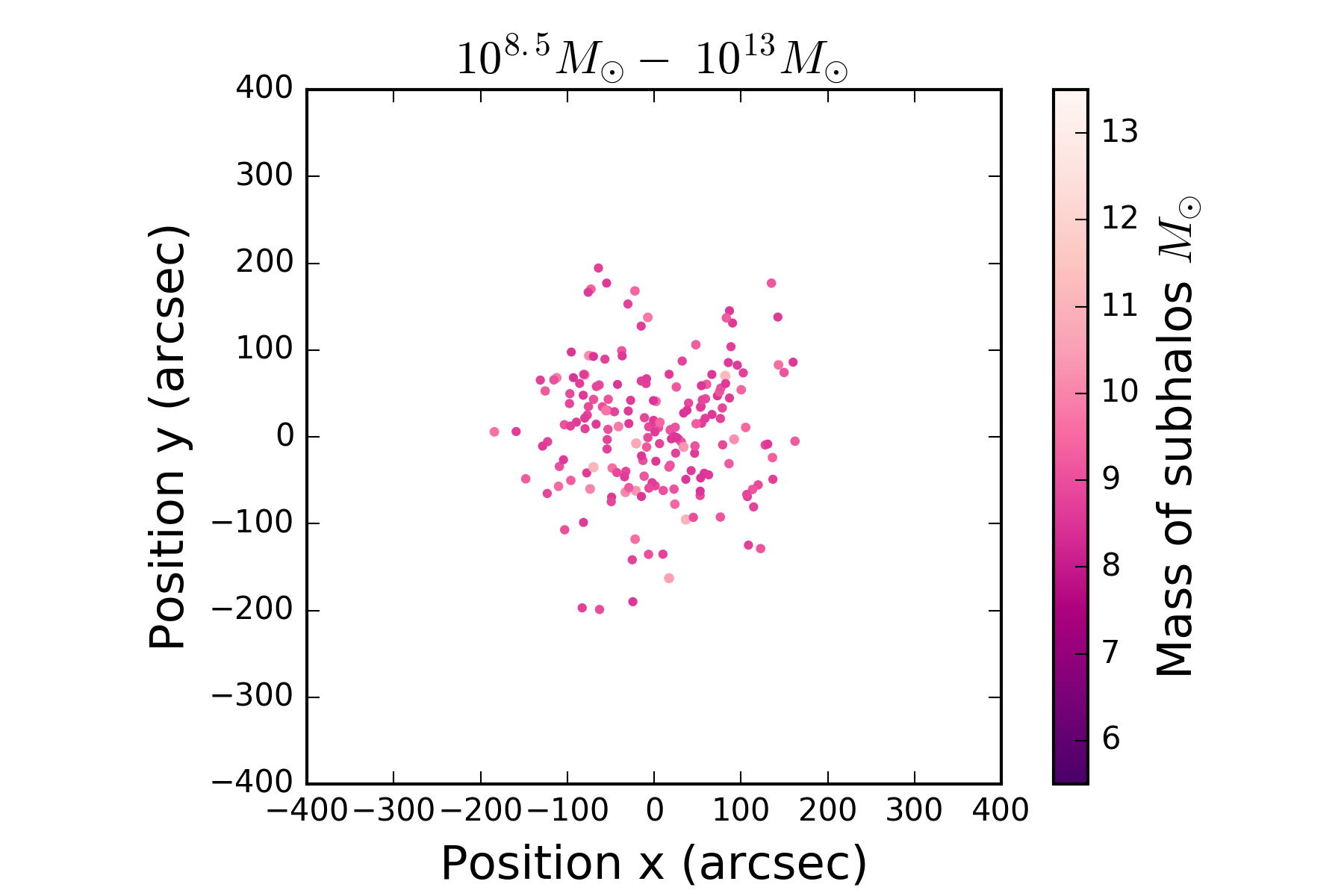}
  
  \end{minipage}
  \hfill
  \begin{minipage}[b]{0.45\textwidth}
    \includegraphics[width=\textwidth]{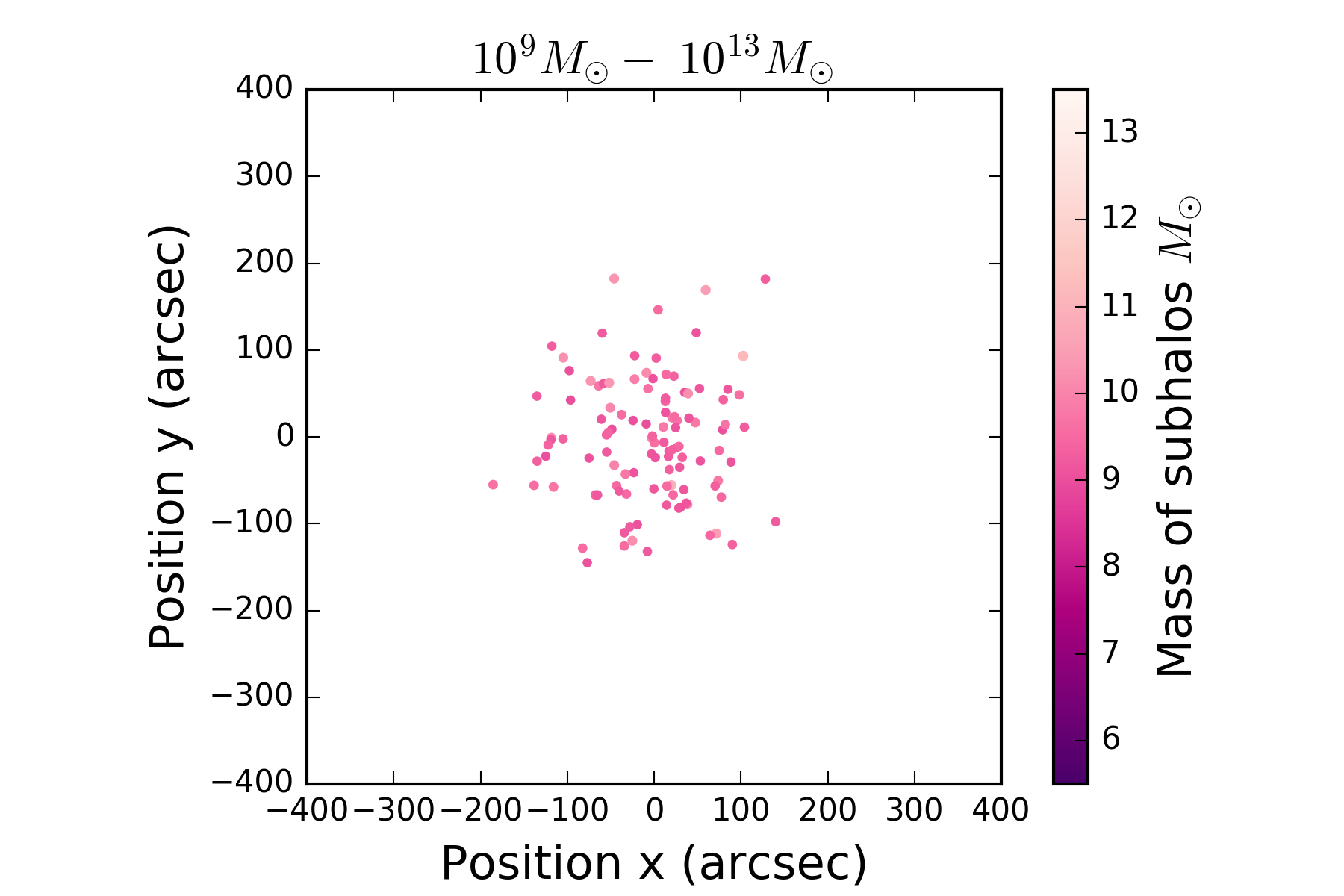}
   
  \end{minipage}

\caption{Distribution of subhalos in different mass ranges around the host according to the Einasto profile. \label{fig:Distributions}} 
  
\end{figure}

Subhalos are randomly distributed around the host galaxy, within a radius $2\times r_{vir}$ where $r_{vir}$ is the virial radius of the host halo according to the Einasto profile with scale radius of 0.81 $r_{vir}$ and a slope $\alpha=1.1$ such that they follow a probability distribution function $P(r)$ of the form.
\begin{equation}
    P(r)\propto r^2 \exp \left\{\frac{-2}{\alpha}\left[\left(\frac{r}{0.81\times r_{vir}}\right)^{\alpha}-1\right]\right\}
    \label{eq:einasto}
\end{equation}
The parameters here are chosen to be consistent with simulations \cite{2010MNRAS.404..502G,1965TrAlm...5...87E,Despali:2016meh,2008MNRAS.391.1685S}.
The distribution of subhalos is shown in figure \ref{fig:Distributions} where the horizontal units are expressed in arcseconds, the conversion to which will be explained below.

We closely follow the image generation parameters as set out in \cite{DiazRivero:2019hxf}. The host and all subhalos are modelled with the Navarro–Frenk–White (NFW) profile. The host halo ($M$) is set to have a fixed concentration parameter of 15 ($c= 15$). The subhalo concentrations ($c$) are obtained using the formula given below

\begin{equation}
    c= \left[0.0279\times\left(\frac{m_{sub}}{10^{16}M_\odot}\right )^{0.36}+1.942\times10^{-5}\times\left(\frac{m_{sub}}{3\times10^{-3}M_\odot}\right)^{0.0879}\right]^{-0.3333}
    \label{eq:conc_sub}
\end{equation}
which is a fit to the results of \cite{2015MNRAS.452.1217C}.
\subsection{Source Modelling}

We use the Sersic Elliptic profile to model the source galaxy behind the Lens 
\begin{equation}
I(R)=I_e\exp\left\{-\left(2n-\frac{1}{3}\right)\left[\left(\frac{R}{R_{sersic}}\right)^{1/n}-1\right]\right\}
\end{equation}
and as in \cite{DiazRivero:2019hxf}, the source is different in each image. We keep the Sersic index $n=1$ while the half-light radius ($R_{sersic}$) is randomly generated in a uniform range
\begin{equation}
    R_{sersic} \sim U[0.1{\rm kpc}, 1{\rm kpc}] 
    \label{eq:rsersic}
\end{equation}
and the components of ellipticity ($e_x, e_y$) are also drawn from a random distribution 
\begin{equation}
    \epsilon_x,\epsilon_y \sim U[-0.5, 0.5]  \text{ with } \sqrt{\epsilon_x^2+\epsilon_y^2} \leq 0.4
    \label{eq:ellipticity}
\end{equation}
coordinates of the center ($x,y$) relative to the central line through the centre of mass of the lenses are drawn from uniform distributions given below.

\begin{equation}
    x,y \sim U[-0.1'', 0.1'']  
    \label{eq:center}
\end{equation}

Having defined the nature of the lens and the source, we then go on to generate the lensed images.

\subsection{Lensed image}

The scale radius $r_s$ and the deflection angle (in arcseconds) for the host halo and subhalos are obtained from \textsc{lenstronomy} by specifying the mass and concentration parameter as described in subsection \ref{subsec:lens}. The scale radius and the deflection angle of the host are fixed due to its constant mass and concentration parameter to 7.4517'' (0.02537 Mpc) and 1.6641'' (0.00567 Mpc) respectively. We assum a source redshift of $z_s = 0.6$ and the lens redshift of $z_l = 0.2$.

To generate the lensed images, we use \textsc{lenstronomy} \cite{Birrer:2018xgm}, which is a python package to model strong gravitational lenses. The images are 79 $\times$79 pixel in size and span a 5''$\times$5'' window with a resolution of 0.06'' as in \cite{DiazRivero:2019hxf}. 

Image model class in \textsc{lenstronomy} create lensed images using the lens and source arguments provided. Additionally, a Gaussian point source function is given to blur the image modelling class. The full width-half maximum of the Gaussian PSF is kept as 0.1''. 
\cite{Birrer:2018xgm,DiazRivero:2019hxf}. 

We generate 100,000 images per class before we attempt to train a CNN to distinguish between them.

\begin{figure}[!tbp]
  \centering
  \begin{minipage}[b]{0.75\textwidth}
    \includegraphics[width=\textwidth]{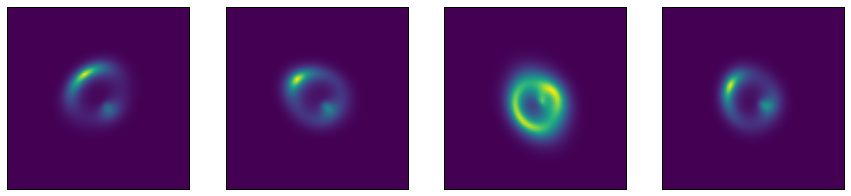}
  
  \end{minipage}
  \vfill
    \centering
  \begin{minipage}[b]{0.75\textwidth}
    \includegraphics[width=\textwidth]{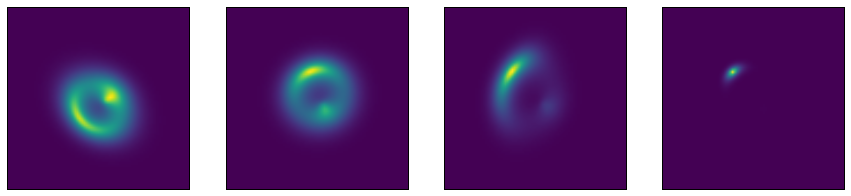}
  
  \end{minipage}
   \vfill
    \centering
  \begin{minipage}[b]{0.75\textwidth}
    \includegraphics[width=\textwidth]{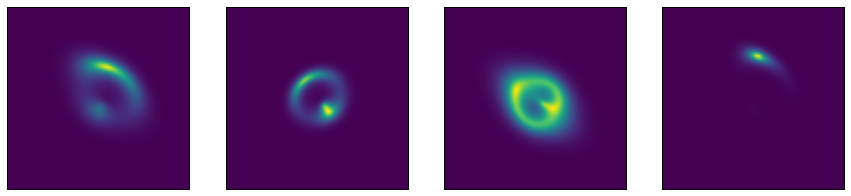}
  
  \end{minipage}
   \vfill
    \centering
  \begin{minipage}[b]{0.75\textwidth}
    \includegraphics[width=\textwidth]{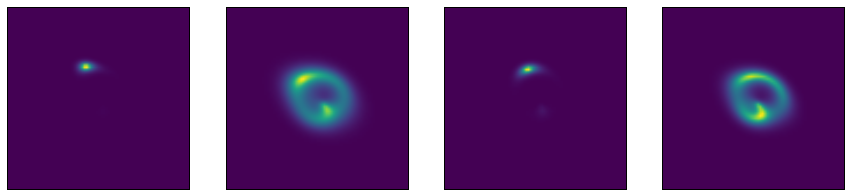}
  
  \end{minipage}
   \vfill
    \centering
  \begin{minipage}[b]{0.75\textwidth}
    \includegraphics[width=\textwidth]{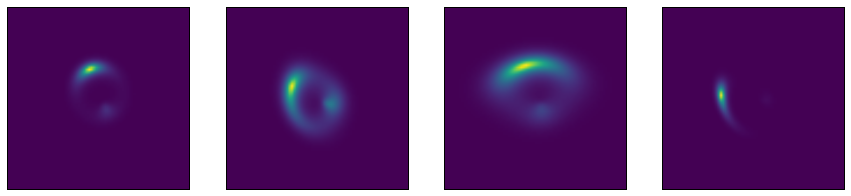}
  
  \end{minipage}
   \vfill
    \centering
  \begin{minipage}[b]{0.75\textwidth}
    \includegraphics[width=\textwidth]{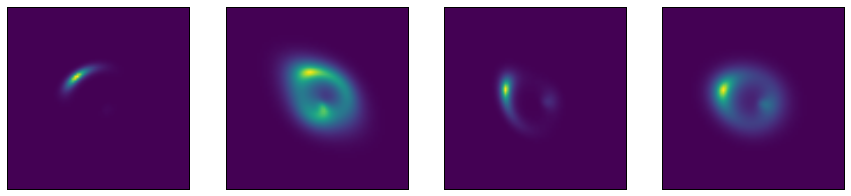}
  
  \end{minipage}
   \vfill
    \centering
  \begin{minipage}[b]{0.75\textwidth}
    \includegraphics[width=\textwidth]{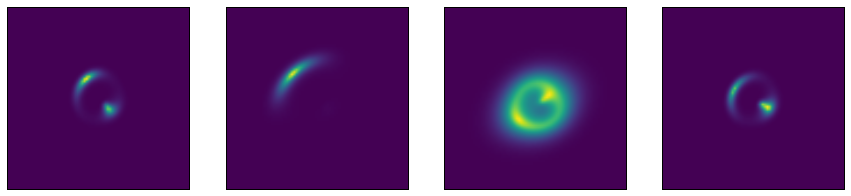}
  
  \end{minipage}
\caption{Strong lensed images generated from \textsc{Lenstronomy} with the data generation method in \ref{sec:data}. Each row corresponds to a different subhalo mass range. For example, first row images have subhalo mass range varying from $10^{9} M_{\odot}- \  10^{13} M_{\odot}$. \label{fig:Images}}

\end{figure}
 \section{Machine Learning: Image recognition}
\label{sec:ml}

Convolutional Neural Networks \cite{10.1162/neco.1989.1.4.541} are extensions of the conventional neural network typically used to process images. Images are convolved through filters the parameters of which which acts as weights in traditional neural networks. As the network progress from each layer, the high-level features of the input images are learned by the network. Non-linear activations are used in the convolutional layer to obtain these feature maps.  The dimensionality of the output of the convolutional layer is reduced by pooling layers. After a series of convolutional (and pooling) layers, a set of fully-connected layers are attached to combine all the features learned and to classify the images.
The weights of each layer are optimized during training via backpropagation. Recent reviews on CNN are \cite{10.1117/12.2512087,article}

We implemented various CNN architectures to classify the images generated from \textsc{lenstronomy}. \textsc{Tensorflow} 2.1 was used to create the model \cite{DBLP:journals/corr/AbadiABBCCCDDDG16}.  Models were trained using \textsc{Nvidia GeForce GTX 1080Ti} and \textsc{RTX 2080Ti} GPUs running \textsc{CUDA} 10.2 \cite{Nickolls:2008:SPP:1365490.1365500}.

The first two models (hereby CNN1 and CNN2) are optimized for a different context \cite{Komiske:2016rsd,Macaluso:2018tck,Moore:2018lsr}. VGG-like architecture and ResNet-like architecture is also adopted in this study. The following subsections briefly explain the architecture and parameters of each network.

\subsubsection*{CNN1}
The network has three convolutional layers and two fully connected layers. The filter size is 64 for all the convolutional layers. The 2D convolution window has a shape of $(8, 8)$ for the first layer and $(4, 4)$ 
for the remaining layers. The weight matrix is initialised with the \textsc{Tensorflow}/\textsc{Keras} default values \cite{chollet2015keras}. Spatial dropout is applied to each convolutional layer to avoid overfitting. A $(2, 2)$ maxpooling is applied to each layer. Two fully connected layers have a size of 128 nodes. 
\begin{figure}[!tbp]
	\centering
    \includestandalone[width=0.7\textwidth]{Architectures/CNN1}
	\caption{Network architecture for CNN1}
	\label{fig:CNN1}
\end{figure}

\subsubsection*{CNN2}
The network has two CNN blocks each with two convolutional layers. The blocks are separated with maxpooling layers of filter size $(2, 2)$. The four convolutional layers have filter sizes 128, 64, 64, 64 respectively. The shape of the filters is kept to be $(4, 4)$.  The blocks are followed by three dense layers with sizes 64, 256 and 256.
\begin{figure}[!tbp]
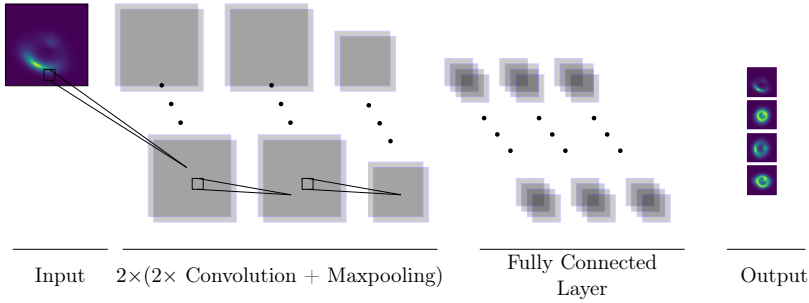

	\centering
    \includestandalone[width=0.7\textwidth]{Architectures/CNN2}
	\caption{Network architecture for CNN2}
	\label{fig:CNN2}
\end{figure}

\subsubsection*{VGG-like}
As the name suggests, the network has a similar structure to the VGG network introduced in \cite{DBLP:journals/corr/SimonyanZ14a}. The network is deep compared to CNN1 and CNN2 and has seven convolutional layers followed by three fully connected layers. The convolutional layers are split into three blocks each consisting of two, two and three convolutional layers. As in  CNN2, blocks are separated by maxpooling layer and spatial dropout.
The filter shape is kept constant $(3, 3)$ throughout the network with size varying in each block. The filter sizes are 64, 128 and 256 respectively for each block. Three dense layers have 4096 nodes.

\begin{figure}[!tbp]
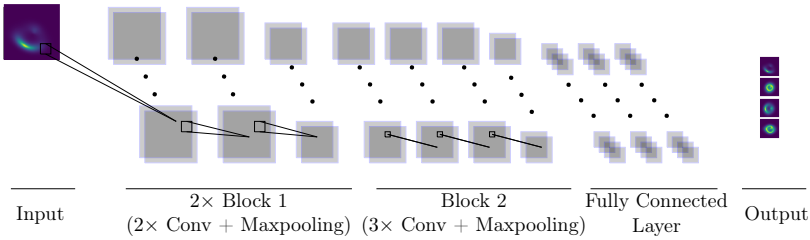

	\centering
    \includestandalone[width=0.7\textwidth]{Architectures/VGG-like}
	\caption{Network architecture for VGG-like network}
	\label{fig:VGG-like}
\end{figure}
\subsubsection*{ResNet-like}
The ``Skip Connection'' presented in \cite{DBLP:journals/corr/HeZRS15} is applied in this network which makes this network ResNet-like in architecture. A ResNet block contains two convolutional layers each followed by Batch Normalization \cite{DBLP:journals/corr/IoffeS15}. The input layer of the block is concatenated to the output layer forming a skip connection.

Ten residual blocks are applied after two convolutional layers of filter size 32 and 64. The convolutional layers in the residual blocks have a filter size of 64. After the residual blocks, there is a convolutional layer with global average pooling. Dropout is applied only to the final dense layer of size 256.

\begin{figure}[!tbp]
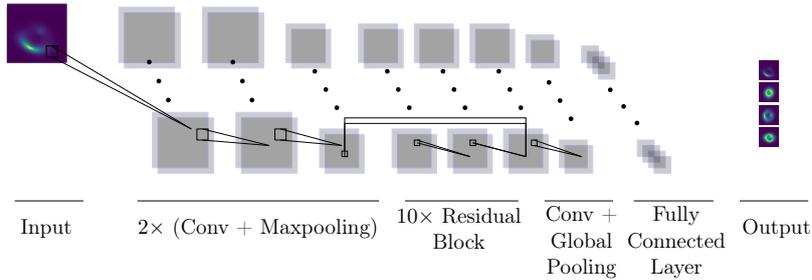

	\centering
    \includestandalone[width=0.7\textwidth]{Architectures/ResNet-like}
	\caption{Network architecture for ResNet-like network}
	\label{fig:VGG-like}
\end{figure}

The activation function used for the inner layers of all the networks is ReLU \cite{Nair:2010:RLU:3104322.3104425} and the output is softmaxed \cite{10.1007/978-3-642-76153-9_28} to obtain probability values. The Adam optimizer \cite{DBLP:journals/corr/KingmaB14} is used with a learning rate of 0.001 in CNN1 and ResNet-like while the learning rate adopted for the VGG-like network is 0.0001. CNN2 is implemented with Adadelta \cite{DBLP:journals/corr/abs-1212-5701} using a learning rate of 0.3. Binary cross-entropy is applied as the loss function for all the networks. The dropout applied in different parts of the above networks have a value of 0.1 (90$\%$ of the image is kept and 10$\%$ is randomly removed to avoid overfitting) \cite{JMLR:v15:srivastava14a}. The networks are trained for 100 epochs with an early patience of 10.

Multi-label classification was performed on these images. 560k images were used for training, 140k images for validation and 70,000 images for testing.

\subsection{Things we tried which didn't work.}

Before moving on to the results, we would like to list some approaches we attempted but which were less successful than what we will present in the next section.

In the early stages of this work, we approached the problem as a multi-class classification with four classes ( $10^6 M_{\odot}-\ 10^{13} M_{\odot}$, $10^7 M_{\odot}-\ 10^{13}$, $10^8 M_{\odot}-\ 10^{13}$,  $10^9 M_{\odot}-\ 10^{13} M_{\odot}$) and considered all the classes to be mutually exclusive.  The average model accuracy was 0.56732, mainly because intermediate ranges were hard to identify using this method - the algorithm could identify the extreme classes at each end of whatever range we chose but not the intermediate classes.

Actually, multiple labels can be associated with each image as the mass ranges are not mutually exclusive.  This is actually to be expected - $10^6 M_{\odot}-\ 10^{13} M_{\odot}$ images contain subhalos of mass $10^7 M_{\odot}$. All mass ranges acts as a subset of $10^6 M_{\odot}-\ 10^{13} M_{\odot}$ mass ranges and all images contains $10^9 M_{\odot}$ subhalos. We therefore decided to switch to multi-label classification where the algorithm doesn't assume the classes are mutually exclusive but merely provides a probability for each.

The multi-label classification then performed on the four classes resulted in an average class-wise accuracy of 0.8764 and average subset accuracy of 0.5601. Even though the class-wise accuracy was on the better side the subset accuracy was not implying a satisfactory performance.  When we went to seven classes, the increased quantity of data to train on resulted in a much better performance.

Binary classification was also applied to this earlier dataset to compare classes with each other, again assuming that the categories were mutually exclusive.  This yielded fairly low accuracies.

To investigate the capability of the networks used, we tried training with Fast Fourier Transforms (FFT) of the images. The networks were trained on grayscale images with pixel intensity being the magnitude of the complex array elements after performing the Fourier transform. We also tried training on colour images by feeding the magnitude and phase as two different colour channels. No significant improvement in the performance was evident with either of these approaches.

\section{Results}
\label{sec:results}
Receiver Operating Characteristics (ROC) curves play a crucial role in measuring the performance of a classifier along with its area-under-curve (AUC). ROC curves are considered to be an important visualising, organising technique to evaluate the performance of a model. The true positive rate (signal efficiency) and false positive rate (background rejection) of a classifier is plotted for each classification threshold to obtain the ROC curves with AUC value evaluating the probability of the classifier in ranking a randomly chosen positive instance higher than a negative instance \cite{FAWCETT2006861,10.1016/S0031-3203(96)00142-2}.  
\begin{table}
\begin{center}
\begin{tabular}{ |c|c|c|c|c|c| } 
 \hline
Sample & CNN1 & CNN2 & VGG-like & ResNet-like  \\ 
 \hline
\ 
$10^6 M_{\odot}-\ 10^{13} M_{\odot}$ & 0.9433 & 0.9565 & 0.9519 & 0.9715  \\ 
$10^{6.5} M_{\odot}-\ 10^{13} M_{\odot}$ & 0.9622 & 0.9737 & 0.9689 & 0.9837   \\ 
$10^7 M_{\odot}-\ 10^{13} M_{\odot}$ & 0.9735 & 0.9831  & 0.9786 & 0.9899\\
$10^{7.5} M_{\odot}-\ 10^{13} M_{\odot}$ & 0.9719 & 0.9814  & 0.9771 & 0.9890   \\ 

$10^8 M_{\odot}-\ 10^{13} M_{\odot}$ & 0.9761  & 0.9849 & 0.9809 & 0.9918\\
$10^{8.5} M_{\odot}-\ 10^{13} M_{\odot}$ & 0.9769 & 0.9859 & 0.9827 & 0.9933   \\ 

 \hline
 \end{tabular}
 
 \caption{The area-under-curve (AUC) values for our multi label classification ROC curves. Larger values of AUC implies better classification and AUC=1 corresponds to perfect classifier. $10^9 M_{\odot}-\ 10^{13} M_{\odot}$ images are labelled as signal according to the multi-labelling of the data, hence acts as a perfect classifier with AUC= 1.\label{table:auctable1}}
\end{center}

\end{table}

 
 


\begin{figure}[!tbp]
  \centering
  \begin{minipage}[b]{0.45\textwidth}
    \includegraphics[width=\textwidth]{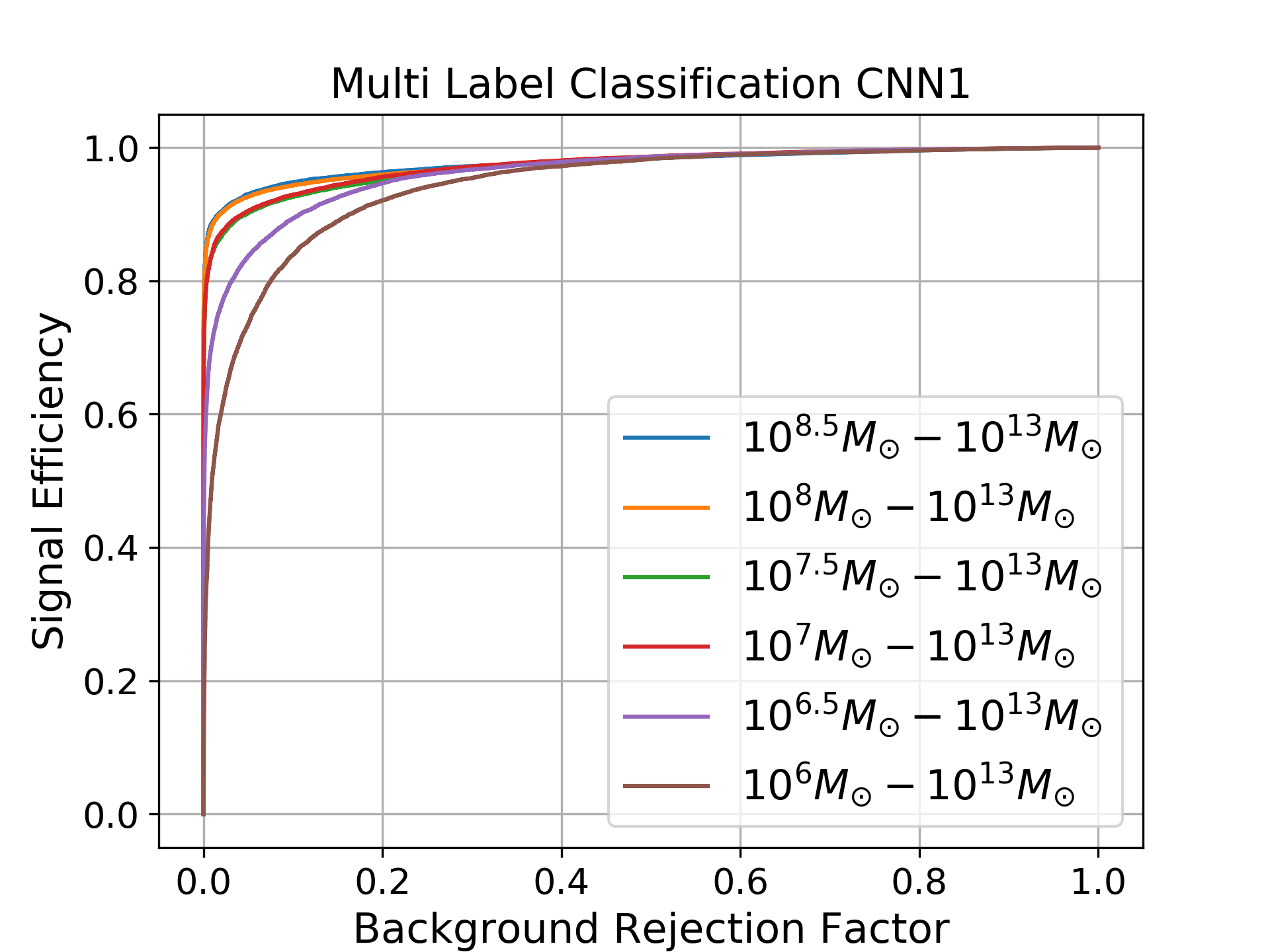}
  
  \end{minipage}
  \hfill
  \begin{minipage}[b]{0.45\textwidth}
    \includegraphics[width=\textwidth]{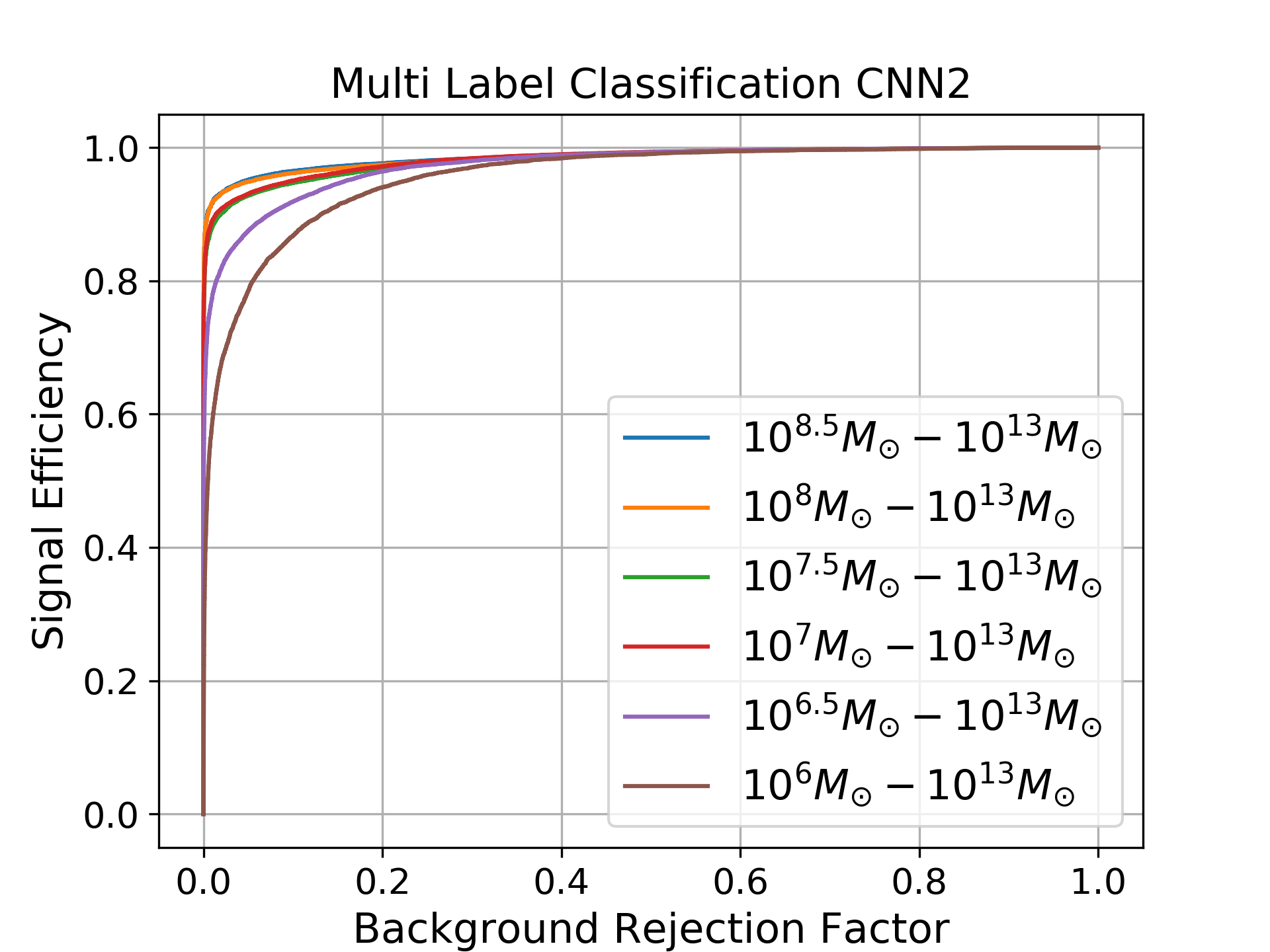}
   
  \end{minipage}

  \begin{minipage}[b]{0.45\textwidth}
    \includegraphics[width=\textwidth]{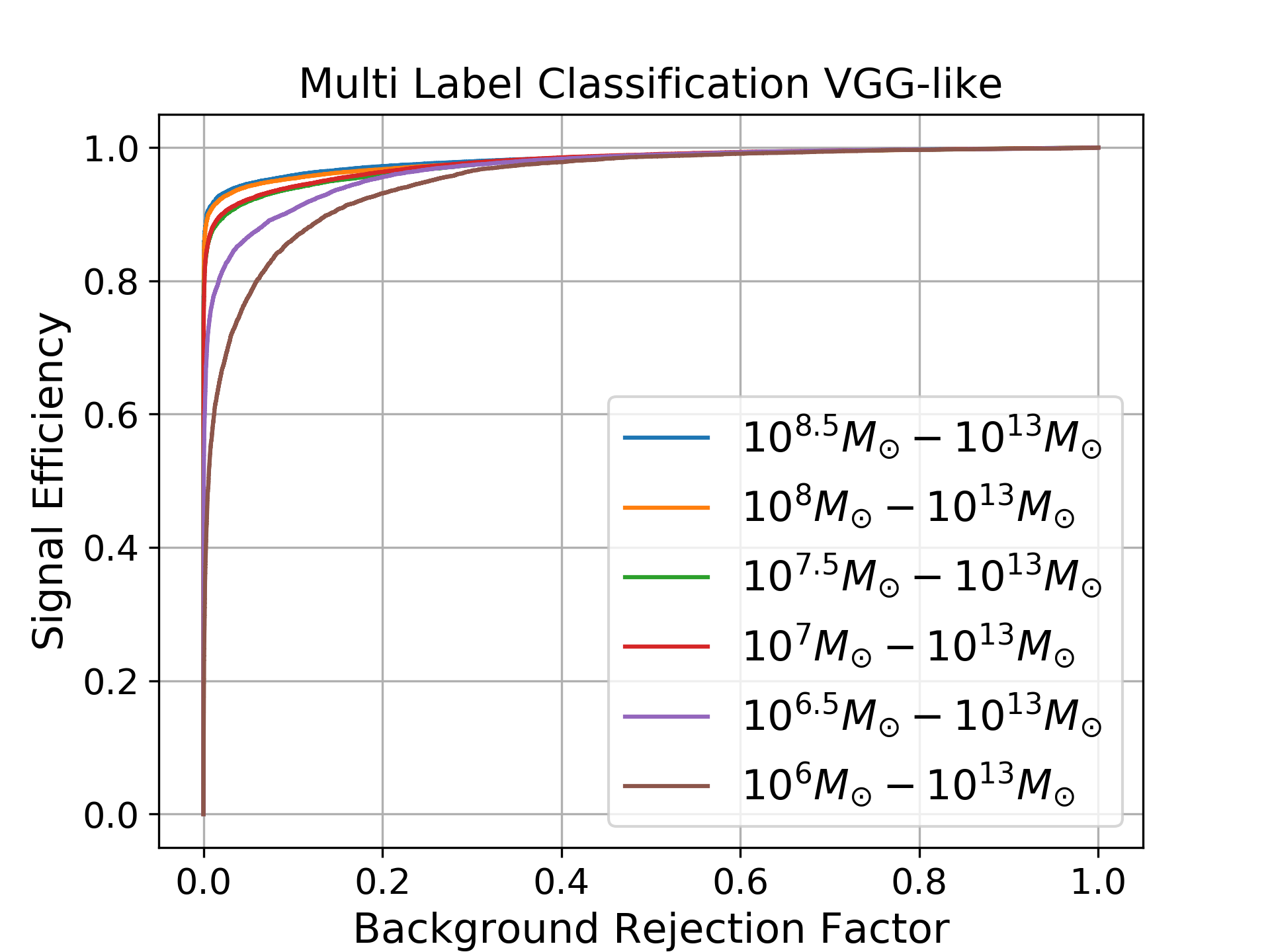}
  
  \end{minipage}
  \hfill
  \begin{minipage}[b]{0.45\textwidth}
    \includegraphics[width=\textwidth]{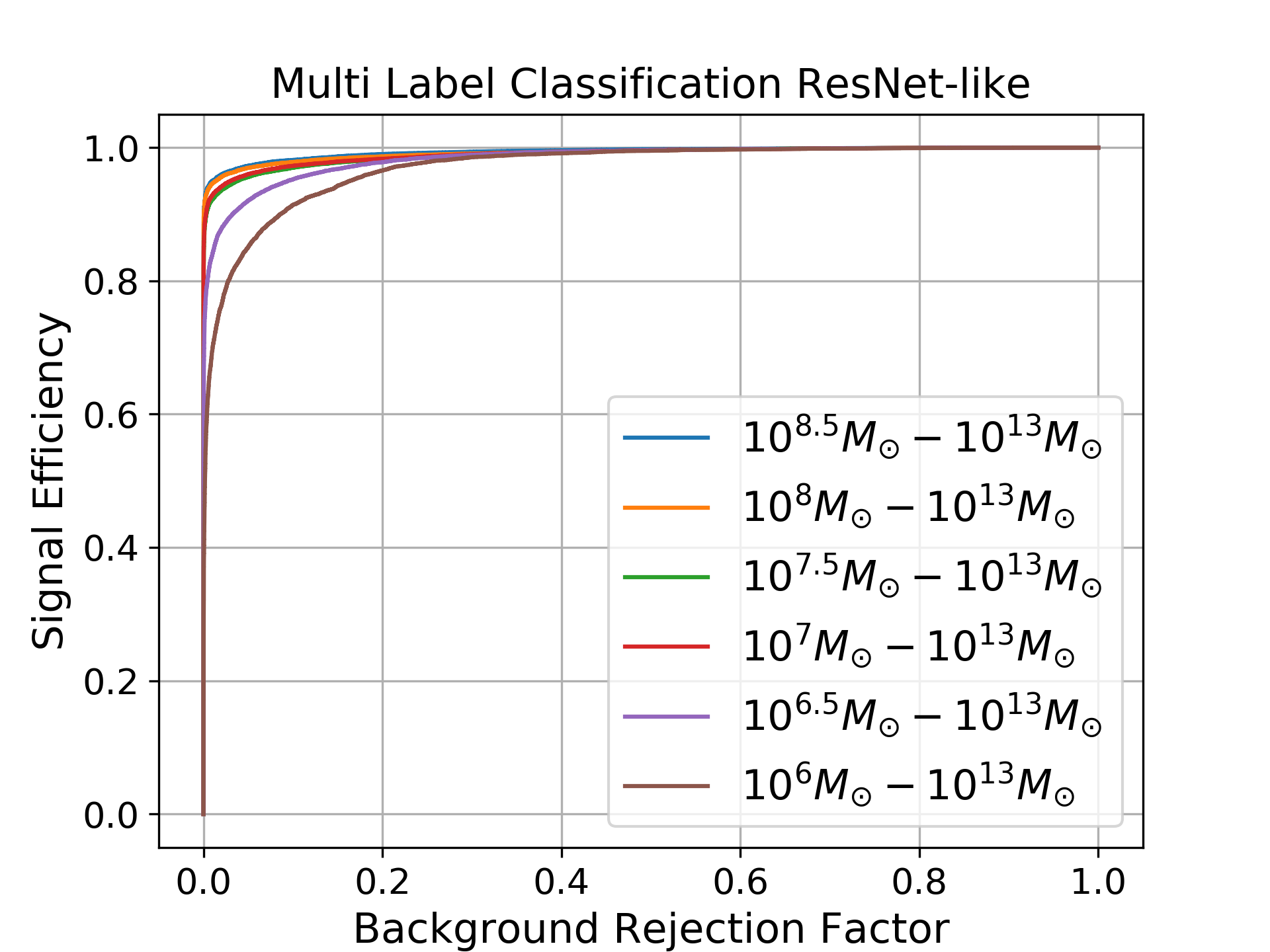}
   
  \end{minipage}

\caption{ROC curves for multi label classification of the lensed images.\label{fig:ML}} 
  
\end{figure}

The ROC curves for our multi-label classifiers are presented in Figure~\ref{fig:ML}. The area-under-curve values are tabulated in Table~\ref{table:auctable1}. The results, as shown in Table~\ref{table:auctable1}, indicate that the networks considered can distinguish lensed images with subhalo masses in the range $10^6 M_{\odot}-\ 10^{13} M_{\odot} $. 

In general, the multi-label classifier can differentiate between each class with an average subset accuracy (when labels match exactly with the predicted values) of 0.7307. The average accuracy value of the model when calculated with individual class label is 0.9509. It is clear from the results that the features learned from the images are the same irrespective of the architecture.

\begin{figure}[!tbp]
  \centering
  
  
  \begin{minipage}[b]{0.5\textwidth}
    \includegraphics[width=\textwidth]{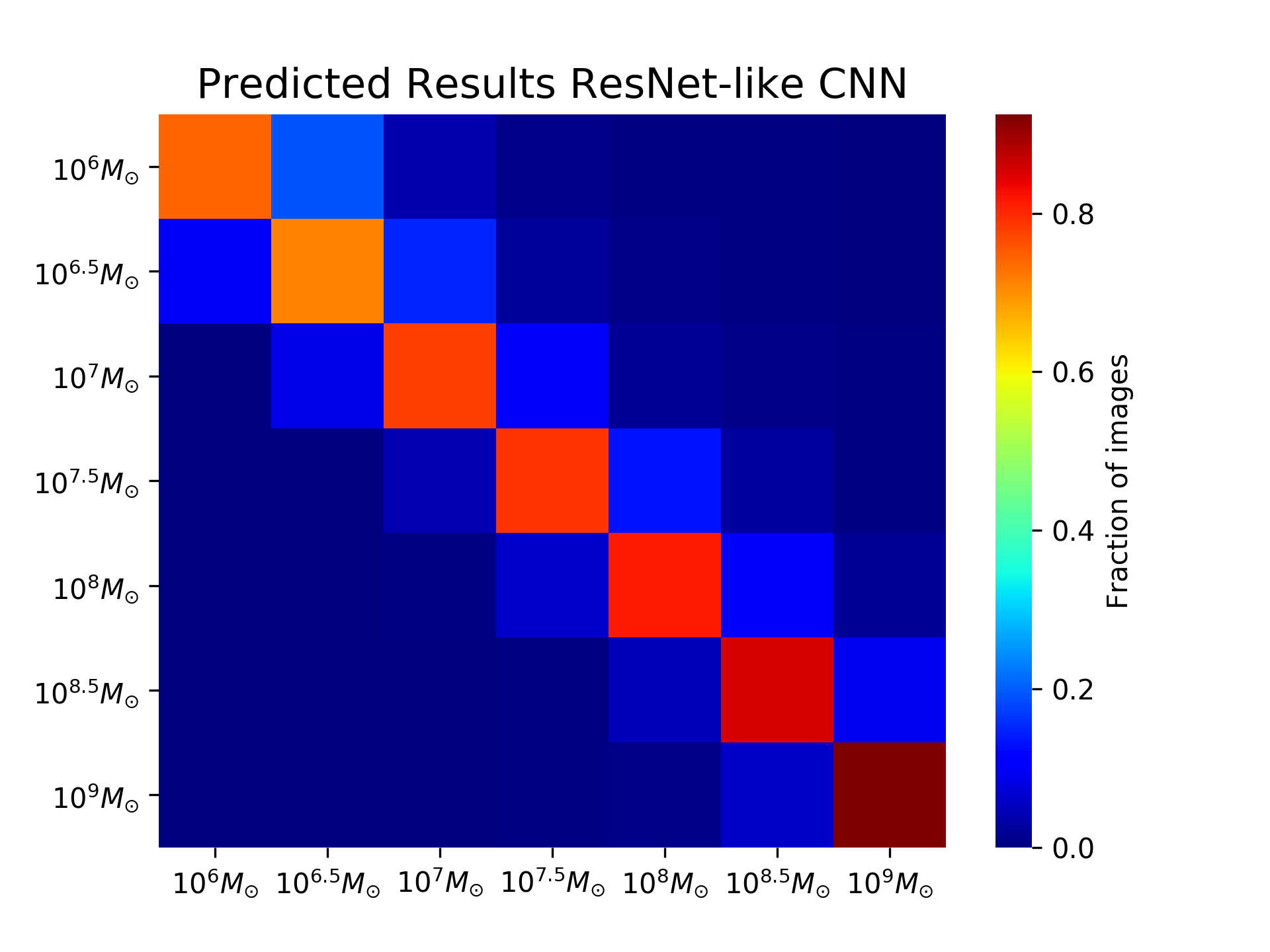}
   
  \end{minipage}
  

\caption{Correlation plot of predicted results from the ResNet- like CNN. 10000 test images are used to predict the probability values of each class. \label{fig:ML-corr}}
  
\end{figure}

\begin{figure}[!tbp]
  \centering
  \begin{minipage}[b]{0.45\textwidth}
    \includegraphics[width=\textwidth]{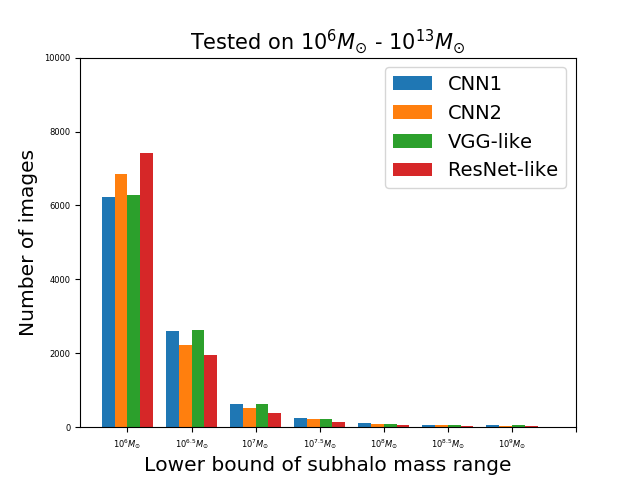}
  
  \end{minipage}
  \hfill
   \begin{minipage}[b]{0.45\textwidth}
    \includegraphics[width=\textwidth]{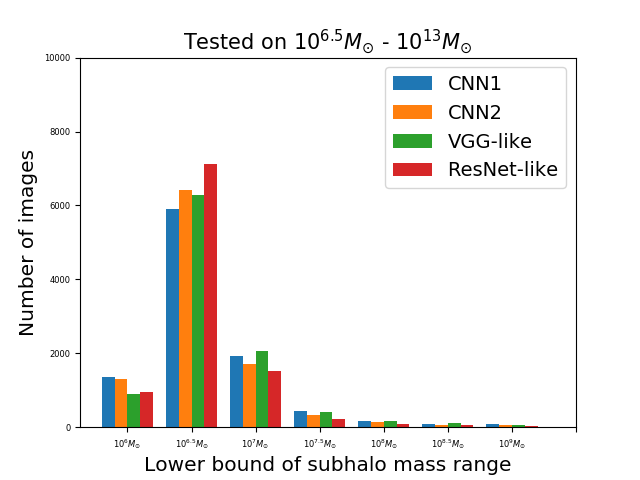}
  
  \end{minipage}
  \hfill
  \begin{minipage}[b]{0.45\textwidth}
    \includegraphics[width=\textwidth]{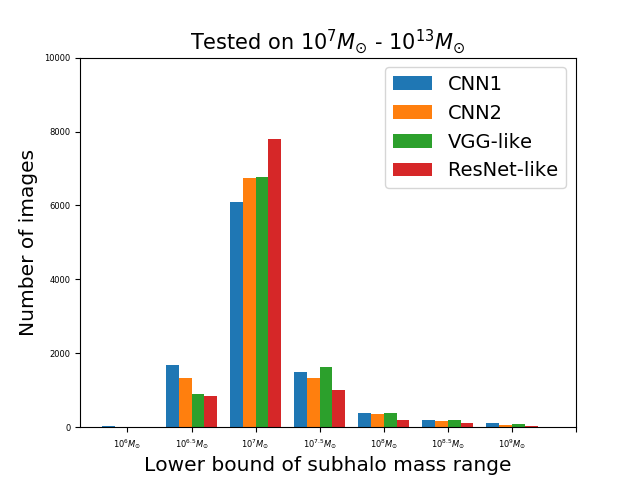}
   
  \end{minipage}
    \hfill
 \begin{minipage}[b]{0.45\textwidth}
    \includegraphics[width=\textwidth]{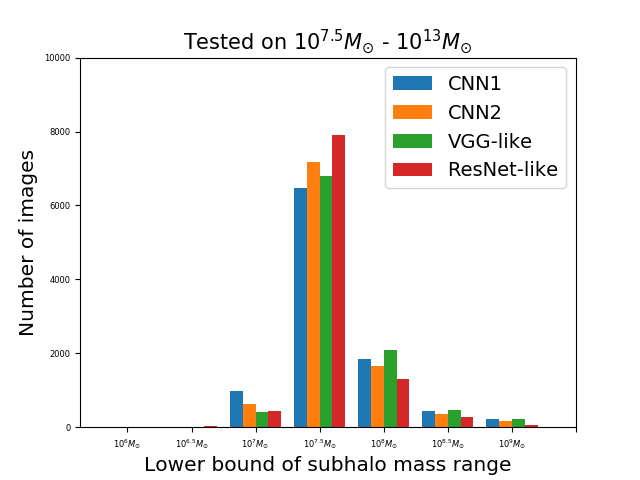}
  
  \end{minipage}
  \hfill
  \begin{minipage}[b]{0.45\textwidth}
    \includegraphics[width=\textwidth]{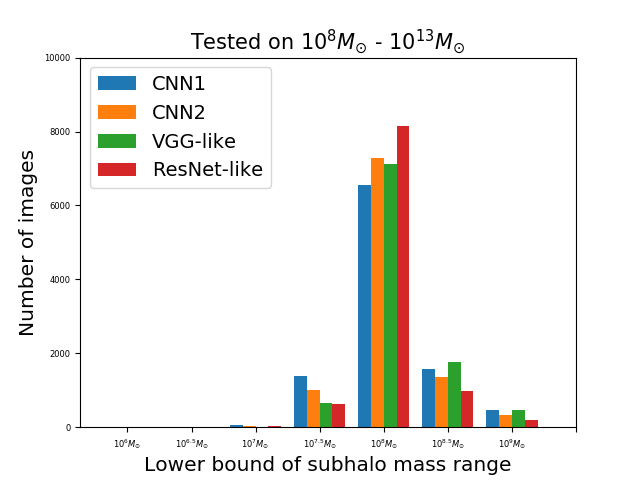}
     \end{minipage}
  \hfill
   \begin{minipage}[b]{0.45\textwidth}
    \includegraphics[width=\textwidth]{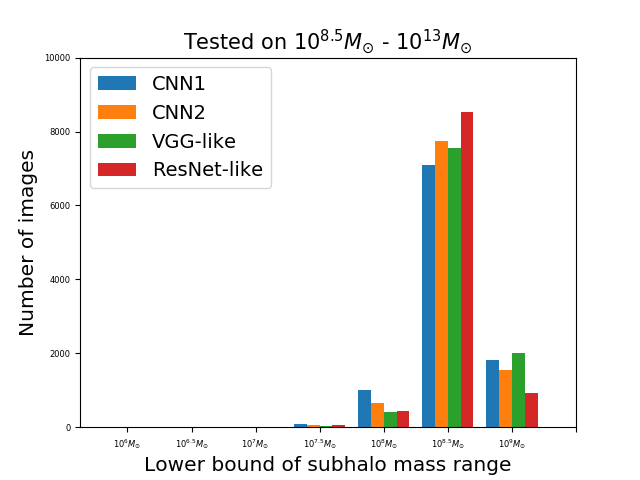}
  
  \end{minipage}
  \hfill
  \begin{minipage}[b]{0.45\textwidth}
    \includegraphics[width=\textwidth]{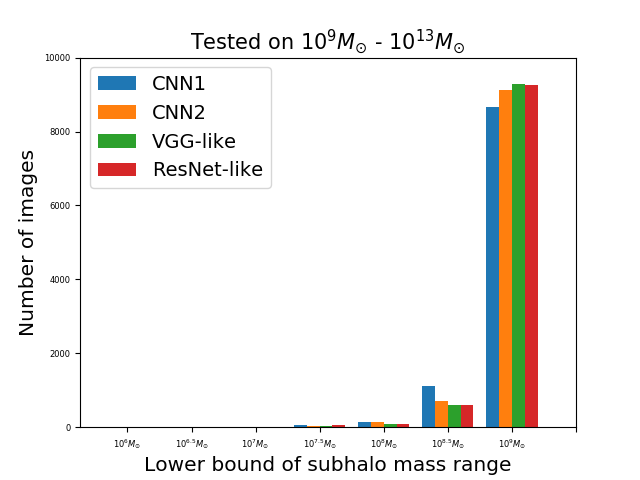}
   
  \end{minipage}

\caption{Distribution of predicted classes of various test image samples (10k images). \label{fig:ML-dist}}
  
\end{figure}

This is very interesting but perhaps detracts from the main physics question we are looking to solve which is how well the algorithm can classify an image into the correct class corresponding to its actual mass.
Figure~\ref{fig:ML-dist} presents the distribution of how the images are being classified into different classes. Each test sample contains 10,000 images. It is clear from the histograms that all networks can identify the features of the images from various classes and usually correctly predict the correct class. The histograms are combined to form the correlation plot shown in Figure~\ref{fig:ML-corr} with diagonal elements representing the fraction of correctly classified images. From Figure~\ref{fig:ML-dist} it is evident that ResNet like CNN identifies the classes more accurately than other three networks. 
Notably, for those images which are misclassified, they are usually placed into adjacent classes, so that it is rare for example for an image generated from a lens containing substructure down to $10^{6.5}M_\odot$ to be misclassified as a halo with a cut-off at $10^{9}M_\odot$, rather it is more likely to be classified as a lens with a $10^{6}M\odot$ or $10^{7}M_\odot$ cut-off.

For example, the fraction of correctly predicted $10^7 M_{\odot}-\ 10^{13} M_{\odot} $ images is 0.7804 but the probability of misclassifying the same as adjacent mass ranges ($10^{6.5} M_{\odot}-\ 10^{13} M_{\odot}$ and $10^{7.5} M_{\odot}-\ 10^{13} M_{\odot}$) are 0.0835 and 0.1004 respectively, meaning that the algorithm predicts that the cut-off lies between $10^{6.5}-10^{7.5}$ with a probability of 0.9643.

\begin{table}
\begin{center}
\begin{tabular}{ |c|c|c| } 
 \hline
Sample & Fraction correctly identified & {Fraction identified within adjacent ranges}   \\ 
 \hline
\ 
$10^6 M_{\odot}-\ 10^{13} M_{\odot}$ & 0.7421 & 0.9364  \\ 
$10^{6.5} M_{\odot}-\ 10^{13} M_{\odot}$ & 0.7119 & 0.9586   \\ 
$10^7 M_{\odot}-\ 10^{13} M_{\odot}$ & 0.7804 & 0.9643 \\ 
$10^{7.5} M_{\odot}-\ 10^{13} M_{\odot}$ & 0.7901 & 0.9634     \\ 

$10^8 M_{\odot}-\ 10^{13} M_{\odot}$ & 0.8138  & 0.9756\\ 
$10^{8.5} M_{\odot}-\ 10^{13} M_{\odot}$ & 0.8536 & 0.9917  \\ 
$10^9 M_{\odot}-\ 10^{13} M_{\odot}$ & 0.9251  & 0.9846 \\ 

 \hline
 \end{tabular}
 
 \caption{The second column represents the fraction of correctly predicted images (for example, what fraction of $10^{7}M_\odot$ images were identified as $10^{7}M_\odot$ images.)  The third column combined the probability of adjacent classes (for example, what fraction of $10^{7}M_\odot$ images were identified as either $10^{6.5}$, $10^{7}$ or $10^{7.5}M_\odot$ images.) This was obtained from the ResNet-like CNN as pictorially represented in Figure~\ref{fig:ML-corr}.}
\end{center}

\end{table}

The images generated following the procedure in Section~\ref{sec:data} are normalized and fed into the CNN without any other preprocessing steps. No data augmentation is being applied while training the networks.

\section{Discussion and Conclusion}
\label{sec:conclusion}
In this work, we have generated simulated images of galaxies which have been lensed by dark matter halos with substructure in order to see if machine learning techniques can tell between different kinds of substructure.  Within the confines of this restricted numerical experiment, we came to the conclusion that they can.  

We considered a spectrum of subhalos within the main lensing halo with distributions that match what is predicted from N-body simulations.  In particular we had seven separate classes of simulated images for each of which the  substructure spectra had different lower mass cut-offs with different minimum masses. These lower cut-offs ranged from $10^6M_\odot$ to $10^9M_\odot$.

We then trained CNN image networks on these simulated lensed images to distinguish between the different classes.  It was shown that we were able to correctly identify the category of cut-off in the vast majority of cases and for those cases when there was a mis-identification, the halo was wrongly identified usually as being in a adjacent class, so that the identification of the cut-off was not so bad.

Ultimately one would hope that by applying machine learning algorithms trained on fake data and then applied to real data that one could identify the lower mass cut-off in the spectrum of sub halos, which would have extremely important implications for the behaviour and ultimate nature of dark matter.  In this paper, we are still quite far from this - we have performed a controlled numerical experiment which for the time being ignores many of the real problems that would be important in such a study.  For example the effects of the non uniform emission of light from both the source and the lens are likely to vary on spatial distances comparable to variations due to the lensing explored here.  In order to quantify whether such effects render such an approach completely unusable when applied to real data, we would have to make use of much more detailed simulations which provided realistically distributed emission of light from baryonic origins in the source and the lens.

We view the current work therefore as an encouraging proof of concept, which could be improved upon and developed further.

There are also further computational avenues which could be investigated in future work, we could use pre-trained CNNs to see if they were more efficient at categorisation and could in principle use GANs to generate more training data less expensively, although we have improved the speed at which images are generated.  We could look at building a model to predict the subhalo mass using ordinal regression and also could launch a comparison between RNN, CNN and cloud network approaches applied to the same problem. We also noted that when we went from four classes to seven classes, the classification power of the algorithm increased, so it would be interesting to experiment with different numbers of classes and images within those classes to see how much accuracy could be improved without overtraining.

Finding the minimum mass of dark matter substructure inside halos may be one of the most important pieces of evidence not only for the existence of dark matter but also for a clue into its particle nature and origin.  It is not clear what approach will end up helping realise this goal.  With this work, we hope to help develop one of those approaches.

  
   
  

  
   

  
   




\acknowledgments
We would like to thank Alexey Boyarsky, David Harvey and  Jorge Penarrubia for useful discussions and valuable comments.
The work of MF was supported partly by the STFC GrantST/L000326/1. SV and MF are also funded by the European Research Council under the European Union’s Horizon 2020 programme (ERC Grant Agreement no.648680 DARK-HORIZONS). SV was the recipient of a Rick Trainor PhD Scholarship at the start of her studies. JF worked extensively on this project as part of King’s Undergraduate Research Fellowship scheme (KURF).



\bibliographystyle{ieeetr}
\bibliography{lens.bib}
\end{document}

%% file: Architectures/CNN1.tex
	\begin{tikzpicture}
		    \node (conv-left) at (0,-0.5){};
		\node (conv-right) at (2,-0.5){};
		\draw[] (conv-left) -- (conv-right);
		\node[align=center] at (1.0,-1){Input};

		\draw (0,2.5) -- (1.5,2.5) -- (1.5,4) -- (0,4) -- (0,2.5);
		
		\node[inner sep=0pt] (input) at (0.7,3.21)
        {\includegraphics[width=1.7cm]{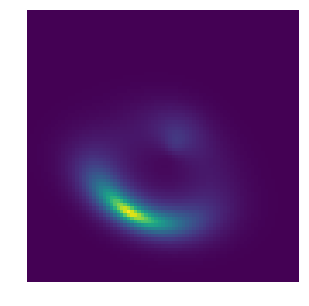}};

		\draw (0.7,2.6) -- (0.9,2.6) -- (0.9,2.8) -- (0.7,2.8) -- (0.7,2.6);
		
		\draw (0.8,2.8) -- (3.3,1.0);
		\draw (0.8,2.6) -- (3.3,1.0);

		
		\draw[fill=black,opacity=0.2,draw=blue] (2.0,2.5) -- (3.5,2.5) -- (3.5,4) -- (2,4) -- (2.0,2.5);
		\draw[fill=black,opacity=0.2,draw=blue] (2.1,2.4) -- (3.6,2.4) -- (3.6,3.9) -- (2.1,3.9) -- (2.1,2.4);

		\node (d-u) at (3.0,2.2){};
		\node (d-l) at (3.1,2){};
		
        \path (d-u) -- (d-l) node [black, font=\Huge, midway,sloped] {$\dots$};
        
		\draw[fill=black,opacity=0.2,draw=blue] (2.6,0.1) -- (4.1,0.1) -- (4.1,1.6) -- (2.6,1.6) -- (2.6,0.1);
		\draw[fill=black,opacity=0.2,draw=blue] (2.7,0) -- (4.2,0) -- (4.2,1.5) -- (2.7,1.5) -- (2.7,0);

 		\draw (3.4,0.6) -- (3.6,0.6) -- (3.6,0.8) -- (3.4,0.8) -- (3.4,0.6);
		
 		\draw (3.5,0.8) -- (5.2,0.5);
 		\draw (3.5,0.6) -- (5.2,0.5);

 		\draw[fill=black,opacity=0.2,draw=blue] (4.0,2.5) -- (5,2.5) -- (5,3.5) -- (4,3.5) -- (4.0,2.5);
 		\draw[fill=black,opacity=0.2,draw=blue] (4.1,2.4) -- (5.1,2.4) -- (5.1,3.4) -- (4.1,3.4) -- (4.1,2.4);

		\node (d-u) at (4.7,2.0){};
		\node (d-l) at (5.0,1.5){};
		
        \path (d-u) -- (d-l) node [black, font=\Huge, midway,sloped] {$\dots$};		
		
 		\draw[fill=black,opacity=0.2,draw=blue] (4.6,0.1) -- (5.6,0.1) -- (5.6,1.1) -- (4.6,1.1) -- (4.6,1.1);
 		\draw[fill=black,opacity=0.2,draw=blue] (4.7,0) -- (5.7,0) -- (5.7,1.0) -- (4.7,1.0) -- (4.7,0);

	    \node (conv-left) at (2,-0.5){};
		\node (conv-right) at (6,-0.5){};
		\draw[] (conv-left) -- (conv-right);
		\node[align=center] at (4.0,-1){$3\times$ (Convolution + \\ Maxpooling)};

		
  		\draw[fill=black,opacity=0.2,draw=blue] (6.0,2.5) -- (6.5,2.5) -- (6.5,3.0) -- (6.0,3.0) -- (6.0,2.5);
  		\draw[fill=black,opacity=0.2,draw=blue] (6.1,2.4) -- (6.6,2.4) -- (6.6,2.9) -- (6.1,2.9) -- (6.1,2.4);
		\draw[fill=black,opacity=0.2,draw=blue] (6.2,2.3) -- (6.7,2.3) -- (6.7,2.8) -- (6.2,2.8) -- (6.2,2.3);
		\draw[fill=black,opacity=0.2,draw=blue] (6.3,2.2) -- (6.8,2.2) -- (6.8,2.7) -- (6.3,2.7) -- (6.3,2.2);

		\node (d-u) at (6.8,1.8){};
		\node (d-l) at (7.2,1.3){};
		
        \path (d-u) -- (d-l) node [black, font=\Huge, midway,sloped] {$\dots$};

 		\draw[fill=black,opacity=0.2,draw=blue] (7.3,0.3) -- (7.8,0.3) -- (7.8,0.8) -- (7.3,0.8) -- (7.3,0.3);
 		\draw[fill=black,opacity=0.2,draw=blue] (7.4,0.2) -- (7.9,0.2) -- (7.9,0.7) -- (7.4,0.7) -- (7.4,0.2);

 		\draw[fill=black,opacity=0.2,draw=blue] (7.5,0.1) -- (8.0,0.1) -- (8.0,0.6) -- (7.5,0.6) -- (7.5,0.1);
 		\draw[fill=black,opacity=0.2,draw=blue] (7.6,0.0) -- (8.1,0.0) -- (8.1,0.5) -- (7.6,0.5) -- (7.6,0.0);


	\draw[fill=black,opacity=0.2,draw=blue] (7.0,2.5) -- (7.5,2.5) -- (7.5,3.0) -- (7.0,3.0) -- (7.0,2.5);
  		\draw[fill=black,opacity=0.2,draw=blue] (7.1,2.4) -- (7.7,2.4) -- (7.7,2.9) -- (7.1,2.9) -- (7.1,2.4);
		\draw[fill=black,opacity=0.2,draw=blue] (7.2,2.3) -- (7.7,2.3) -- (7.7,2.8) -- (7.2,2.8) -- (7.2,2.3);
		\draw[fill=black,opacity=0.2,draw=blue] (7.3,2.2) -- (7.8,2.2) -- (7.8,2.7) -- (7.3,2.7) -- (7.3,2.2);
		
		\node (d-u) at (7.8,1.8){};
		\node (d-l) at (8.2,1.3){};
		
        \path (d-u) -- (d-l) node [black, font=\Huge, midway,sloped] {$\dots$};

 		\draw[fill=black,opacity=0.2,draw=blue] (8.3,0.3) -- (8.8,0.3) -- (8.8,0.8) -- (8.3,0.8) -- (8.3,0.3);
 		\draw[fill=black,opacity=0.2,draw=blue] (8.4,0.2) -- (8.9,0.2) -- (8.9,0.7) -- (8.4,0.7) -- (8.4,0.2);

 		\draw[fill=black,opacity=0.2,draw=blue] (8.5,0.1) -- (9.0,0.1) -- (9.0,0.6) -- (8.5,0.6) -- (8.5,0.1);
 		\draw[fill=black,opacity=0.2,draw=blue] (8.6,0.0) -- (9.1,0.0) -- (9.1,0.5) -- (8.6,0.5) -- (8.6,0.0);

	    \node (conv-left) at (6.5,-0.5){};
		\node (conv-right) at (10,-0.5){};
		\draw[] (conv-left) -- (conv-right);
		\node[align=center] at (8.5,-1){Fully Connected \\Layer};

  		\draw[fill=black,opacity=0.2,draw=blue] (10.5,2.3) -- (11,2.3) -- (11,2.8) -- (10.5,2.8) -- (10.5,2.3);

  		\draw[fill=black,opacity=0.2,draw=blue] (10.5,1.7) -- (11,1.7) -- (11,2.2) -- (10.5,2.2) -- (10.5,1.7);

  		\draw[fill=black,opacity=0.2,draw=blue] (10.5,1.1) -- (11,1.1) -- (11,1.6) -- (10.5,1.6) -- (10.5,1.1);

 		\draw[fill=black,opacity=0.2,draw=blue] (10.5,0.5) -- (11,0.5) -- (11,1.0) -- (10.5,1.0) -- (10.5,1.0);  		

     \node[inner sep=0pt] (im1e09) at (10.75,2.55)
     {\includegraphics[width=.6cm]{Architectures/im_1e9.png}};

     \node[inner sep=0pt] (im1e08) at (10.75,1.95)
    {\includegraphics[width=.6cm]{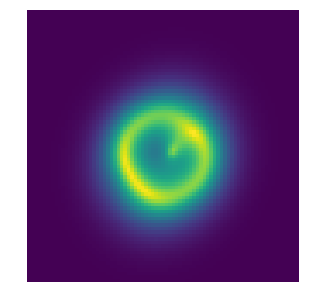}};
   
    \node[inner sep=0pt] (im1e07) at (10.75,1.35)
    {\includegraphics[width=.6cm]{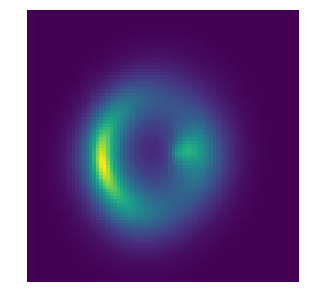}};

    \node[inner sep=0pt] (im1e06) at (10.75,0.75)
    {\includegraphics[width=.6cm]{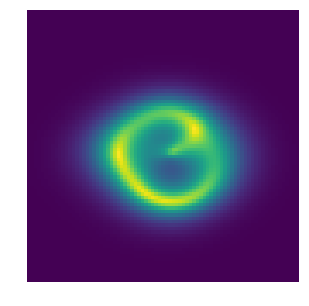}};

 	    \node (conv-left) at (10.,-0.5){};
 		\node (conv-right) at (11.7,-0.5){};
 		\draw[] (conv-left) -- (conv-right);
 		\node[align=center] at (11,-1){Output};

	\end{tikzpicture}

%% file: Architectures/CNN2.tex
	\begin{tikzpicture}
		    \node (conv-left) at (0,-0.5){};
		\node (conv-right) at (2,-0.5){};
		\draw[] (conv-left) -- (conv-right);
		\node[align=center] at (1.0,-1){Input};

				\node[inner sep=0pt] (russell) at (0.7,3.21)
    {\includegraphics[width=1.7cm]{Architectures/im_1e9.png}};

		\draw (0,2.5) -- (1.5,2.5) -- (1.5,4) -- (0,4) -- (0,2.5);
		
		\draw (0.7,2.6) -- (0.9,2.6) -- (0.9,2.8) -- (0.7,2.8) -- (0.7,2.6);
		
		\draw (0.8,2.8) -- (3.3,1.0);
		\draw (0.8,2.6) -- (3.3,1.0);

		
		\draw[fill=black,opacity=0.2,draw=blue] (2.0,2.5) -- (3.5,2.5) -- (3.5,4) -- (2,4) -- (2.0,2.5);
		\draw[fill=black,opacity=0.2,draw=blue] (2.1,2.4) -- (3.6,2.4) -- (3.6,3.9) -- (2.1,3.9) -- (2.1,2.4);

		\node (d-u) at (3.0,2.2){};
		\node (d-l) at (3.1,2){};
		
        \path (d-u) -- (d-l) node [black, font=\Huge, midway,sloped] {$\dots$};

		\draw[fill=black,opacity=0.2,draw=blue] (2.6,0.1) -- (4.1,0.1) -- (4.1,1.6) -- (2.6,1.6) -- (2.6,0.1);
		\draw[fill=black,opacity=0.2,draw=blue] (2.7,0) -- (4.2,0) -- (4.2,1.5) -- (2.7,1.5) -- (2.7,0);

 		\draw (3.4,0.6) -- (3.6,0.6) -- (3.6,0.8) -- (3.4,0.8) -- (3.4,0.6);
		
 		\draw (3.5,0.8) -- (5.2,0.5);
 		\draw (3.5,0.6) -- (5.2,0.5);


		\draw[fill=black,opacity=0.2,draw=blue] (4.0,2.5) -- (5.5,2.5) -- (5.5,4) -- (4,4) -- (4,2.5);
		\draw[fill=black,opacity=0.2,draw=blue] (4.1,2.4) -- (5.6,2.4) -- (5.6,3.9) -- (4.1,3.9) -- (4.1,2.4);

		\node (d-u) at (5.0,2.2){};
		\node (d-l) at (5.1,2){};
		
        \path (d-u) -- (d-l) node [black, font=\Huge, midway,sloped] {$\dots$};

		\draw[fill=black,opacity=0.2,draw=blue] (4.6,0.1) -- (6.1,0.1) -- (6.1,1.6) -- (4.6,1.6) -- (4.6,0.1);
		\draw[fill=black,opacity=0.2,draw=blue] (4.7,0) -- (6.2,0) -- (6.2,1.5) -- (4.7,1.5) -- (4.7,0);

		\draw (5.4,0.6) -- (5.6,0.6) -- (5.6,0.8) -- (5.4,0.8) -- (5.4,0.6);
		
 		\draw (5.5,0.8) -- (7.2,0.5);
 		\draw (5.5,0.6) -- (7.2,0.5);
		
 		\draw[fill=black,opacity=0.2,draw=blue] (6.0,2.5) -- (7,2.5) -- (7,3.5) -- (6,3.5) -- (6.0,2.5);
 		\draw[fill=black,opacity=0.2,draw=blue] (6.1,2.4) -- (7.1,2.4) -- (7.1,3.4) -- (6.1,3.4) -- (6.1,2.4);

		\node (d-u) at (6.7,2.0){};
		\node (d-l) at (7.0,1.5){};
		
        \path (d-u) -- (d-l) node [black, font=\Huge, midway,sloped] {$\dots$};		
		
 		\draw[fill=black,opacity=0.2,draw=blue] (6.6,0.1) -- (7.6,0.1) -- (7.6,1.1) -- (6.6,1.1) -- (6.6,1.1);
 		\draw[fill=black,opacity=0.2,draw=blue] (6.7,0) -- (7.7,0) -- (7.7,1.0) -- (6.7,1.0) -- (6.7,0);

	    \node (conv-left) at (2,-0.5){};
		\node (conv-right) at (8,-0.5){};
		\draw[] (conv-left) -- (conv-right);
		\node[align=center] at (5,-1){$2\times$(2$\times$ Convolution + Maxpooling)};

		
  		\draw[fill=black,opacity=0.2,draw=blue] (8.0,2.5) -- (8.5,2.5) -- (8.5,3.0) -- (8.0,3.0) -- (8.0,2.5);
  		\draw[fill=black,opacity=0.2,draw=blue] (8.1,2.4) -- (8.6,2.4) -- (8.6,2.9) -- (8.1,2.9) -- (8.1,2.4);
		\draw[fill=black,opacity=0.2,draw=blue] (8.2,2.3) -- (8.7,2.3) -- (8.7,2.8) -- (8.2,2.8) -- (8.2,2.3);
		\draw[fill=black,opacity=0.2,draw=blue] (8.3,2.2) -- (8.8,2.2) -- (8.8,2.7) -- (8.3,2.7) -- (8.3,2.2);

		\node (d-u) at (8.8,1.8){};
		\node (d-l) at (9.2,1.3){};
		
        \path (d-u) -- (d-l) node [black, font=\Huge, midway,sloped] {$\dots$};

 		\draw[fill=black,opacity=0.2,draw=blue] (9.3,0.3) -- (9.8,0.3) -- (9.8,0.8) -- (9.3,0.8) -- (9.3,0.3);
 		\draw[fill=black,opacity=0.2,draw=blue] (9.4,0.2) -- (9.9,0.2) -- (9.9,0.7) -- (9.4,0.7) -- (9.4,0.2);

 		\draw[fill=black,opacity=0.2,draw=blue] (9.5,0.1) -- (10.0,0.1) -- (10.0,0.6) -- (9.5,0.6) -- (9.5,0.1);
 		\draw[fill=black,opacity=0.2,draw=blue] (9.6,0.0) -- (10.1,0.0) -- (10.1,0.5) -- (9.6,0.5) -- (9.6,0.0);


	\draw[fill=black,opacity=0.2,draw=blue] (9.0,2.5) -- (9.5,2.5) -- (9.5,3.0) -- (9.0,3.0) -- (9.0,2.5);
  		\draw[fill=black,opacity=0.2,draw=blue] (9.1,2.4) -- (9.7,2.4) -- (9.7,2.9) -- (9.1,2.9) -- (9.1,2.4);
		\draw[fill=black,opacity=0.2,draw=blue] (9.2,2.3) -- (9.7,2.3) -- (9.7,2.8) -- (9.2,2.8) -- (9.2,2.3);
		\draw[fill=black,opacity=0.2,draw=blue] (9.3,2.2) -- (9.8,2.2) -- (9.8,2.7) -- (9.3,2.7) -- (9.3,2.2);
		
		\node (d-u) at (9.8,1.8){};
		\node (d-l) at (10.2,1.3){};
		
        \path (d-u) -- (d-l) node [black, font=\Huge, midway,sloped] {$\dots$};

 		\draw[fill=black,opacity=0.2,draw=blue] (10.3,0.3) -- (10.8,0.3) -- (10.8,0.8) -- (10.3,0.8) -- (10.3,0.3);
 		\draw[fill=black,opacity=0.2,draw=blue] (10.4,0.2) -- (10.9,0.2) -- (10.9,0.7) -- (10.4,0.7) -- (10.4,0.2);

 		\draw[fill=black,opacity=0.2,draw=blue] (10.5,0.1) -- (11.0,0.1) -- (11.0,0.6) -- (10.5,0.6) -- (10.5,0.1);
 		\draw[fill=black,opacity=0.2,draw=blue] (10.6,0.0) -- (11.1,0.0) -- (11.1,0.5) -- (10.6,0.5) -- (10.6,0.0);


	\draw[fill=black,opacity=0.2,draw=blue] (10.0,2.5) -- (10.5,2.5) -- (10.5,3.0) -- (10.0,3.0) -- (10.0,2.5);
  		\draw[fill=black,opacity=0.2,draw=blue] (10.1,2.4) -- (10.7,2.4) -- (10.7,2.9) -- (10.1,2.9) -- (10.1,2.4);
		\draw[fill=black,opacity=0.2,draw=blue] (10.2,2.3) -- (10.7,2.3) -- (10.7,2.8) -- (10.2,2.8) -- (10.2,2.3);
		\draw[fill=black,opacity=0.2,draw=blue] (10.3,2.2) -- (10.8,2.2) -- (10.8,2.7) -- (10.3,2.7) -- (10.3,2.2);
		
		\node (d-u) at (10.8,1.8){};
		\node (d-l) at (11.2,1.3){};
		
        \path (d-u) -- (d-l) node [black, font=\Huge, midway,sloped] {$\dots$};

 		\draw[fill=black,opacity=0.2,draw=blue] (11.3,0.3) -- (11.8,0.3) -- (11.8,0.8) -- (11.3,0.8) -- (11.3,0.3);
 		\draw[fill=black,opacity=0.2,draw=blue] (11.4,0.2) -- (11.9,0.2) -- (11.9,0.7) -- (11.4,0.7) -- (11.4,0.2);

 		\draw[fill=black,opacity=0.2,draw=blue] (11.5,0.1) -- (12.0,0.1) -- (12.0,0.6) -- (11.5,0.6) -- (11.5,0.1);
 		\draw[fill=black,opacity=0.2,draw=blue] (11.6,0.0) -- (12.1,0.0) -- (12.1,0.5) -- (11.6,0.5) -- (11.6,0.0);

	    \node (conv-left) at (8.5,-0.5){};
		\node (conv-right) at (12.5,-0.5){};
		\draw[] (conv-left) -- (conv-right);
		\node[align=center] at (10.5,-1){Fully Connected \\Layer};
		
  		\draw[fill=black,opacity=0.2,draw=blue] (13.5,2.3) -- (14,2.3) -- (14,2.8) -- (13.5,2.8) -- (13.5,2.3);		
		
  		\draw[fill=black,opacity=0.2,draw=blue] (13.5,1.7) -- (14,1.7) -- (14,2.2) -- (13.5,2.2) -- (13.5,1.7);
  		
  		\draw[fill=black,opacity=0.2,draw=blue] (13.5,1.1) -- (14,1.1) -- (14,1.6) -- (13.5,1.6) -- (13.5,1.1);  		

 		\draw[fill=black,opacity=0.2,draw=blue] (13.5,0.5) -- (14,0.5) -- (14,1.0) -- (13.5,1.0) -- (13.5,1.0);  		

    \node[inner sep=0pt] (im1e09) at (13.75,2.55)
    {\includegraphics[width=.6cm]{Architectures/im_1e9.png}};

     \node[inner sep=0pt] (im1e08) at (13.75,1.95)
    {\includegraphics[width=.6cm]{Architectures/im_1e8.png}};
   
    \node[inner sep=0pt] (im1e07) at (13.75,1.35)
    {\includegraphics[width=.6cm]{Architectures/im_1e7.png}};

    \node[inner sep=0pt] (im1e06) at (13.75,0.75)
    {\includegraphics[width=.6cm]{Architectures/im_1e6.png}};

	    \node (conv-left) at (13.,-0.5){};
		\node (conv-right) at (14.7,-0.5){};
		\draw[] (conv-left) -- (conv-right);
		\node[align=center] at (14,-1){Output};

	\end{tikzpicture}

%% file: Architectures/VGG-like.tex
	\begin{tikzpicture}
		    \node (conv-left) at (0,-0.5){};
		\node (conv-right) at (1.5,-0.5){};
		\draw[] (conv-left) -- (conv-right);
		\node[align=center] at (0.7,-1){Input};

		\draw (0,2) -- (1,2) -- (1,3) -- (0,3) -- (0,2);
		
		\node[inner sep=0pt] (input) at (0.5,2.5)
        {\includegraphics[width=1.2cm]{Architectures/im_1e9.png}};		
		
		\draw (0.7,2.1) -- (0.9,2.1) -- (0.9,2.3) -- (0.7,2.3) -- (0.7,2.1);
		
		\draw (0.8,2.3) -- (3.3,0.8);
		\draw (0.8,2.1) -- (3.3,0.8);

		
 		\draw[fill=black,opacity=0.2,draw=blue] (2.0,2) -- (3,2) -- (3,3) -- (2,3) -- (2.0,2);
 		\draw[fill=black,opacity=0.2,draw=blue] (2.1,1.9) -- (3.1,1.9) -- (3.1,2.9) -- (2.1,2.9) -- (2.1,1.9);

 		\node (d-u) at (2.7,1.7){};
 		\node (d-l) at (2.8,1.5){};
		
         \path (d-u) -- (d-l) node [black, font=\Huge, midway,sloped] {$\dots$};

 		\draw[fill=black,opacity=0.2,draw=blue] (2.6,0.1) -- (3.6,0.1) -- (3.6,1.1) -- (2.6,1.1) -- (2.6,0.1);
 		\draw[fill=black,opacity=0.2,draw=blue] (2.7,0) -- (3.7,0) -- (3.7,1.) -- (2.7,1.) -- (2.7,0);

  		\draw (3.4,0.6) -- (3.6,0.6) -- (3.6,0.8) -- (3.4,0.8) -- (3.4,0.6);
		
  		\draw (3.5,0.8) -- (4.7,0.5);
  		\draw (3.5,0.6) -- (4.7,0.5);

		
 		\draw[fill=black,opacity=0.2,draw=blue] (3.5,2) -- (4.5,2) -- (4.5,3) -- (3.5,3) -- (3.5,2);
 		\draw[fill=black,opacity=0.2,draw=blue] (3.6,1.9) -- (4.6,1.9) -- (4.6,2.9) -- (3.6,2.9) -- (3.6,1.9);

 		\node (d-u) at (4.2,1.7){};
 		\node (d-l) at (4.3,1.5){};
		
         \path (d-u) -- (d-l) node [black, font=\Huge, midway,sloped] {$\dots$};

 		\draw[fill=black,opacity=0.2,draw=blue] (4.1,0.1) -- (5.1,0.1) -- (5.1,1.1) -- (4.1,1.1) -- (4.1,0.1);
 		\draw[fill=black,opacity=0.2,draw=blue] (4.2,0) -- (5.2,0) -- (5.2,1.) -- (4.2,1.) -- (4.2,0);

  		\draw (4.8,0.6) -- (5,0.6) -- (5,0.8) -- (4.8,0.8) -- (4.8,0.6);
		
  		\draw (4.9,0.8) -- (6,0.5);
  		\draw (4.9,0.6) -- (6,0.5);

 		\draw[fill=black,opacity=0.2,draw=blue] (5,2) -- (5.7,2) -- (5.7,2.7) -- (5,2.7) -- (5,2);
 		\draw[fill=black,opacity=0.2,draw=blue] (5.1,1.9) -- (5.8,1.9) -- (5.8,2.6) -- (5.1,2.6) -- (5.1,1.9);

 		\node (d-u) at (5.6,1.7){};
 		\node (d-l) at (5.9,1.2){};
		
         \path (d-u) -- (d-l) node [black, font=\Huge, midway,sloped] {$\dots$};		
		
  		\draw[fill=black,opacity=0.2,draw=blue] (5.6,0.1) -- (6.3,0.1) -- (6.3,0.8) -- (5.6,0.8) -- (5.6,0.1);
  		\draw[fill=black,opacity=0.2,draw=blue] (5.7,0) -- (6.4,0) -- (6.4,0.7) -- (5.7,0.7) -- (5.7,0);

 	    \node (conv-left) at (2.2,-0.5){};
 		\node (conv-right) at (6.8,-0.5){};
 		\draw[] (conv-left) -- (conv-right);
 		\node[align=center] at (4.5,-1){$2\times$ Block 1 \\ ($2 \times$ Conv + Maxpooling)};


 		\draw[fill=black,opacity=0.2,draw=blue] (6.3,2) -- (7,2) -- (7,2.7) -- (6.3,2.7) -- (6.3,2);
 		\draw[fill=black,opacity=0.2,draw=blue] (6.4,1.9) -- (7.1,1.9) -- (7.1,2.6) -- (6.4,2.6) -- (6.4,1.9);

    	\node (d-u) at (6.9,1.7){};
 		\node (d-l) at (7.2,1.2){};
		
         \path (d-u) -- (d-l) node [black, font=\Huge, midway,sloped] {$\dots$};		
		
  		\draw[fill=black,opacity=0.2,draw=blue] (6.9,0.1) -- (7.6,0.1) -- (7.6,0.8) -- (6.9,0.8) -- (6.9,0.1);
  		\draw[fill=black,opacity=0.2,draw=blue] (7.0,0) -- (7.7,0) -- (7.7,0.7) -- (7,0.7) -- (7,0);

	
		\draw[fill=black,opacity=0.2,draw=blue] (7.3,2) -- (8,2) -- (8,2.7) -- (7.3,2.7) -- (7.3,2);
 		\draw[fill=black,opacity=0.2,draw=blue] (7.4,1.9) -- (8.1,1.9) -- (8.1,2.6) -- (7.4,2.6) -- (7.4,1.9);

 		\node (d-u) at (7.9,1.7){};
 		\node (d-l) at (8.2,1.2){};
		
         \path (d-u) -- (d-l) node [black, font=\Huge, midway,sloped] {$\dots$};

  		\draw[fill=black,opacity=0.2,draw=blue] (7.9,0.1) -- (8.6,0.1) -- (8.6,0.8) -- (7.9,0.8) -- (7.9,0.1);
  		\draw[fill=black,opacity=0.2,draw=blue] (8.0,0) -- (8.7,0) -- (8.7,0.7) -- (8,0.7) -- (8,0);

 		\draw (8.3,0.5) -- (8.4,0.5) -- (8.4,0.6) -- (8.3,0.6) -- (8.3,0.5);
		
  		\draw (8.35,0.55) -- (9.3,0.3);
  		\draw (8.35,0.55) -- (9.3,0.3);


 		\draw[fill=black,opacity=0.2,draw=blue] (8.3,2) -- (9,2) -- (9,2.7) -- (8.3,2.7) -- (8.3,2);
 		\draw[fill=black,opacity=0.2,draw=blue] (8.4,1.9) -- (9.1,1.9) -- (9.1,2.6) -- (8.4,2.6) -- (8.4,1.9);

 		\node (d-u) at (8.9,1.7){};
 		\node (d-l) at (9.2,1.2){};
		
         \path (d-u) -- (d-l) node [black, font=\Huge, midway,sloped] {$\dots$};		
		
  		\draw[fill=black,opacity=0.2,draw=blue] (8.9,0.1) -- (9.6,0.1) -- (9.6,0.8) -- (8.9,0.8) -- (8.9,0.1);
  		\draw[fill=black,opacity=0.2,draw=blue] (9.0,0) -- (9.7,0) -- (9.7,0.7) -- (9,0.7) -- (9,0);

 		\draw (9.3,0.5) -- (9.4,0.5) -- (9.4,0.6) -- (9.3,0.6) -- (9.3,0.5);
		
  		\draw (9.35,0.55) -- (10.3,0.3);
  		\draw (9.35,0.55) -- (10.3,0.3);

 		\draw[fill=black,opacity=0.2,draw=blue] (9.3,2) -- (9.8,2) -- (9.8,2.5) -- (9.3,2.5) -- (9.3,2);
 		\draw[fill=black,opacity=0.2,draw=blue] (9.4,1.9) -- (9.9,1.9) -- (9.9,2.4) -- (9.4,2.4) -- (9.4,1.9);

  		\node (d-u) at (9.8,1.5){};
  		\node (d-l) at (10.1,1.0){};
		
          \path (d-u) -- (d-l) node [black, font=\Huge, midway,sloped] {$\dots$};		
		
  		\draw[fill=black,opacity=0.2,draw=blue] (9.9,0.1) -- (10.4,0.1) -- (10.4,0.6) -- (9.9,0.6) -- (9.9,0.1);
  		\draw[fill=black,opacity=0.2,draw=blue] (10,0) -- (10.5,0) -- (10.5,0.5) -- (10,0.5) -- (10,0);

  		\draw (7.3,0.5) -- (7.4,0.5) -- (7.4,0.6) -- (7.3,0.6) -- (7.3,0.5);
		
  		\draw (7.35,0.55) -- (8.3,0.3);
  		\draw (7.35,0.55) -- (8.3,0.3);

 	    \node (conv-left) at (7.,-0.5){};
 		\node (conv-right) at (11.0,-0.5){};
 		\draw[] (conv-left) -- (conv-right);
 		\node[align=center] at (9.,-1){ Block 2 \\ ($3 \times$ Conv + Maxpooling)};

		
  		\draw[fill=black,opacity=0.2,draw=blue] (10.3,2) -- (10.6,2) -- (10.6,2.3) -- (10.3,2.3) -- (10.3,2.0);
  		\draw[fill=black,opacity=0.2,draw=blue] (10.4,1.9) -- (10.7,1.9) -- (10.7,2.2) -- (10.4,2.2) -- (10.4,1.9);
  		\draw[fill=black,opacity=0.2,draw=blue] (10.5,1.8) -- (10.8,1.8) -- (10.8,2.1) -- (10.5,2.1) -- (10.5,1.8);
  		\draw[fill=black,opacity=0.2,draw=blue] (10.6,1.7) -- (10.9,1.7) -- (10.9,2.) -- (10.6,2.) -- (10.6,1.7);

  		\node (d-u) at (10.9,1.4){};
  		\node (d-l) at (11.3,0.9){};
		
          \path (d-u) -- (d-l) node [black, font=\Huge, midway,sloped] {$\dots$};

  		\draw[fill=black,opacity=0.2,draw=blue] (11.3,0.3) -- (11.6,0.3) -- (11.6,0.6) -- (11.3,0.6) -- (11.3,0.3);
  		\draw[fill=black,opacity=0.2,draw=blue] (11.4,0.2) -- (11.7,0.2) -- (11.7,0.5) -- (11.4,0.5) -- (11.4,0.2);

  		\draw[fill=black,opacity=0.2,draw=blue] (11.5,0.1) -- (11.8,0.1) -- (11.8,0.4) -- (11.5,0.4) -- (11.5,0.1);
  		\draw[fill=black,opacity=0.2,draw=blue] (11.6,0.0) -- (11.9,0.0) -- (11.9,0.3) -- (11.6,0.3) -- (11.6,0.0);

		
  		\draw[fill=black,opacity=0.2,draw=blue] (11,2) -- (11.3,2) -- (11.3,2.3) -- (11,2.3) -- (11,2.0);
  		\draw[fill=black,opacity=0.2,draw=blue] (11.1,1.9) -- (11.4,1.9) -- (11.4,2.2) -- (11.1,2.2) -- (11.1,1.9);
  		\draw[fill=black,opacity=0.2,draw=blue] (11.2,1.8) -- (11.5,1.8) -- (11.5,2.1) -- (11.2,2.1) -- (11.2,1.8);
  		\draw[fill=black,opacity=0.2,draw=blue] (11.3,1.7) -- (11.6,1.7) -- (11.6,2) -- (11.3,2) -- (11.3,1.7);

  		\node (d-u) at (11.6,1.4){};
  		\node (d-l) at (12,0.9){};
		
          \path (d-u) -- (d-l) node [black, font=\Huge, midway,sloped] {$\dots$};

  		\draw[fill=black,opacity=0.2,draw=blue] (12.,0.3) -- (12.3,0.3) -- (12.3,0.6) -- (12.,0.6) -- (12.,0.3);
  		\draw[fill=black,opacity=0.2,draw=blue] (12.1,0.2) -- (12.4,0.2) -- (12.4,0.5) -- (12.1,0.5) -- (12.1,0.2);

  		\draw[fill=black,opacity=0.2,draw=blue] (12.2,0.1) -- (12.5,0.1) -- (12.5,0.4) -- (12.2,0.4) -- (12.2,0.1);
  		\draw[fill=black,opacity=0.2,draw=blue] (12.3,0.0) -- (12.6,0.0) -- (12.6,0.3) -- (12.3,0.3) -- (12.3,0.0);

					
  		\draw[fill=black,opacity=0.2,draw=blue] (11.7,2) -- (12,2) -- (12,2.3) -- (11.7,2.3) -- (11.7,2.0);
  		\draw[fill=black,opacity=0.2,draw=blue] (11.8,1.9) -- (12.1,1.9) -- (12.1,2.2) -- (11.8,2.2) -- (11.8,1.9);
  		\draw[fill=black,opacity=0.2,draw=blue] (11.9,1.8) -- (12.2,1.8) -- (12.2,2.1) -- (11.9,2.1) -- (11.9,1.8);
  		\draw[fill=black,opacity=0.2,draw=blue] (12.,1.7) -- (12.3,1.7) -- (12.3,2) -- (12.,2) -- (12.,1.7);

  		\node (d-u) at (12.3,1.4){};
  		\node (d-l) at (12.7,0.9){};
		
          \path (d-u) -- (d-l) node [black, font=\Huge, midway,sloped] {$\dots$};

  		\draw[fill=black,opacity=0.2,draw=blue] (12.7,0.3) -- (13,0.3) -- (13,0.6) -- (12.7,0.6) -- (12.7,0.3);
  		\draw[fill=black,opacity=0.2,draw=blue] (12.8,0.2) -- (13.1,0.2) -- (13.1,0.5) -- (12.8,0.5) -- (12.8,0.2);

  		\draw[fill=black,opacity=0.2,draw=blue] (12.9,0.1) -- (13.2,0.1) -- (13.2,0.4) -- (12.9,0.4) -- (12.9,0.1);
  		\draw[fill=black,opacity=0.2,draw=blue] (13.,0.0) -- (13.3,0.0) -- (13.3,0.3) -- (13.,0.3) -- (13.,0.0);

 	    \node (conv-left) at (11.1,-0.5){};
 		\node (conv-right) at (13.8,-0.5){};
 		\draw[] (conv-left) -- (conv-right);
 		\node[align=center] at (12.5,-1){Fully Connected \\Layer};

   		\draw[fill=black,opacity=0.2,draw=blue] (14.5,1.7) -- (14.8,1.7) -- (14.8,2) -- (14.5,2) -- (14.5,1.7);		
   		\draw[fill=black,opacity=0.2,draw=blue] (14.5,1.3) -- (14.8,1.3) -- (14.8,1.6) -- (14.5,1.6) -- (14.5,1.3);		
		
   		\draw[fill=black,opacity=0.2,draw=blue] (14.5,0.9) -- (14.8,0.9) -- (14.8,1.2) -- (14.5,1.2) -- (14.5,0.9);		

   		\draw[fill=black,opacity=0.2,draw=blue] (14.5,0.5) -- (14.8,0.5) -- (14.8,0.8) -- (14.5,0.8) -- (14.5,0.5);		

    \node[inner sep=0pt] (im1e09) at (14.65,1.85)
    {\includegraphics[width=.4cm]{Architectures/im_1e9.png}};

     \node[inner sep=0pt] (im1e08) at (14.65,1.45)
    {\includegraphics[width=.4cm]{Architectures/im_1e8.png}};
   
    \node[inner sep=0pt] (im1e07) at (14.65,1.05)
    {\includegraphics[width=.4cm]{Architectures/im_1e7.png}};

     \node[inner sep=0pt] (im1e06) at (14.65,0.65)
     {\includegraphics[width=.4cm]{Architectures/im_1e6.png}};

 	    \node (conv-left) at (14.,-0.5){};
 		\node (conv-right) at (15.5,-0.5){};
 		\draw[] (conv-left) -- (conv-right);
 		\node[align=center] at (14.8,-1){Output};

	\end{tikzpicture}

%% file: Architectures/ResNet-like.tex
	\begin{tikzpicture}
		    \node (conv-left) at (0,-0.5){};
		\node (conv-right) at (1.5,-0.5){};
		\draw[] (conv-left) -- (conv-right);
		\node[align=center] at (0.7,-1){Input};

		\draw (0,2) -- (1,2) -- (1,3) -- (0,3) -- (0,2);
		
		\node[inner sep=0pt] (input) at (0.5,2.5)
        {\includegraphics[width=1.2cm]{Architectures/im_1e9.png}};		
		
		\draw (0.7,2.1) -- (0.9,2.1) -- (0.9,2.3) -- (0.7,2.3) -- (0.7,2.1);
		
		\draw (0.8,2.3) -- (3.3,0.8);
		\draw (0.8,2.1) -- (3.3,0.8);

		
 		\draw[fill=black,opacity=0.2,draw=blue] (2.0,2) -- (3,2) -- (3,3) -- (2,3) -- (2.0,2);
 		\draw[fill=black,opacity=0.2,draw=blue] (2.1,1.9) -- (3.1,1.9) -- (3.1,2.9) -- (2.1,2.9) -- (2.1,1.9);

 		\node (d-u) at (2.7,1.7){};
 		\node (d-l) at (2.8,1.5){};
		
         \path (d-u) -- (d-l) node [black, font=\Huge, midway,sloped] {$\dots$};

 		\draw[fill=black,opacity=0.2,draw=blue] (2.6,0.1) -- (3.6,0.1) -- (3.6,1.1) -- (2.6,1.1) -- (2.6,0.1);
 		\draw[fill=black,opacity=0.2,draw=blue] (2.7,0) -- (3.7,0) -- (3.7,1.) -- (2.7,1.) -- (2.7,0);

  		\draw (3.4,0.6) -- (3.6,0.6) -- (3.6,0.8) -- (3.4,0.8) -- (3.4,0.6);
		
  		\draw (3.5,0.8) -- (4.7,0.5);
  		\draw (3.5,0.6) -- (4.7,0.5);

		
 		\draw[fill=black,opacity=0.2,draw=blue] (3.5,2) -- (4.5,2) -- (4.5,3) -- (3.5,3) -- (3.5,2);
 		\draw[fill=black,opacity=0.2,draw=blue] (3.6,1.9) -- (4.6,1.9) -- (4.6,2.9) -- (3.6,2.9) -- (3.6,1.9);

 		\node (d-u) at (4.2,1.7){};
 		\node (d-l) at (4.3,1.5){};
		
         \path (d-u) -- (d-l) node [black, font=\Huge, midway,sloped] {$\dots$};

 		\draw[fill=black,opacity=0.2,draw=blue] (4.1,0.1) -- (5.1,0.1) -- (5.1,1.1) -- (4.1,1.1) -- (4.1,0.1);
 		\draw[fill=black,opacity=0.2,draw=blue] (4.2,0) -- (5.2,0) -- (5.2,1.) -- (4.2,1.) -- (4.2,0);

  		\draw (4.8,0.6) -- (5,0.6) -- (5,0.8) -- (4.8,0.8) -- (4.8,0.6);
		
  		\draw (4.9,0.8) -- (6,0.5);
  		\draw (4.9,0.6) -- (6,0.5);

 		\draw[fill=black,opacity=0.2,draw=blue] (5,2) -- (5.7,2) -- (5.7,2.7) -- (5,2.7) -- (5,2);
 		\draw[fill=black,opacity=0.2,draw=blue] (5.1,1.9) -- (5.8,1.9) -- (5.8,2.6) -- (5.1,2.6) -- (5.1,1.9);

 		\node (d-u) at (5.6,1.7){};
 		\node (d-l) at (5.9,1.2){};
		
         \path (d-u) -- (d-l) node [black, font=\Huge, midway,sloped] {$\dots$};		
		
  		\draw[fill=black,opacity=0.2,draw=blue] (5.6,0.1) -- (6.3,0.1) -- (6.3,0.8) -- (5.6,0.8) -- (5.6,0.1);
  		\draw[fill=black,opacity=0.2,draw=blue] (5.7,0) -- (6.4,0) -- (6.4,0.7) -- (5.7,0.7) -- (5.7,0);

 		\draw (6,0.3) -- (6.1,0.3) -- (6.1,0.4) -- (6,0.4) -- (6,0.3);

    	\draw (6.05,0.4) -- (6.05,1.0) -- (9.3,1.0) -- (9.3,0.3);%
    	\draw (6.05,0.3) -- (6.05,0.9) -- (9.3,0.9) -- (9.3,0.3);%

 	    \node (conv-left) at (2.2,-0.5){};
 		\node (conv-right) at (6.8,-0.5){};
 		\draw[] (conv-left) -- (conv-right);
 		\node[align=center] at (4.5,-1){ $2 \times$ (Conv + Maxpooling)};


 		\draw[fill=black,opacity=0.2,draw=blue] (6.3,2) -- (7,2) -- (7,2.7) -- (6.3,2.7) -- (6.3,2);
 		\draw[fill=black,opacity=0.2,draw=blue] (6.4,1.9) -- (7.1,1.9) -- (7.1,2.6) -- (6.4,2.6) -- (6.4,1.9);

    	\node (d-u) at (6.9,1.7){};
 		\node (d-l) at (7.2,1.2){};
		
         \path (d-u) -- (d-l) node [black, font=\Huge, midway,sloped] {$\dots$};		
		
  		\draw[fill=black,opacity=0.2,draw=blue] (6.9,0.1) -- (7.6,0.1) -- (7.6,0.8) -- (6.9,0.8) -- (6.9,0.1);
  		\draw[fill=black,opacity=0.2,draw=blue] (7.0,0) -- (7.7,0) -- (7.7,0.7) -- (7,0.7) -- (7,0);

	
		\draw[fill=black,opacity=0.2,draw=blue] (7.3,2) -- (8,2) -- (8,2.7) -- (7.3,2.7) -- (7.3,2);
 		\draw[fill=black,opacity=0.2,draw=blue] (7.4,1.9) -- (8.1,1.9) -- (8.1,2.6) -- (7.4,2.6) -- (7.4,1.9);

 		\node (d-u) at (7.9,1.7){};
 		\node (d-l) at (8.2,1.2){};
		
         \path (d-u) -- (d-l) node [black, font=\Huge, midway,sloped] {$\dots$};

  		\draw[fill=black,opacity=0.2,draw=blue] (7.9,0.1) -- (8.6,0.1) -- (8.6,0.8) -- (7.9,0.8) -- (7.9,0.1);
  		\draw[fill=black,opacity=0.2,draw=blue] (8.0,0) -- (8.7,0) -- (8.7,0.7) -- (8,0.7) -- (8,0);

 		\draw (8.3,0.5) -- (8.4,0.5) -- (8.4,0.6) -- (8.3,0.6) -- (8.3,0.5);
		
  		\draw (8.35,0.55) -- (9.3,0.3);
  		\draw (8.35,0.55) -- (9.3,0.3);


 		\draw[fill=black,opacity=0.2,draw=blue] (8.3,2) -- (9,2) -- (9,2.7) -- (8.3,2.7) -- (8.3,2);
 		\draw[fill=black,opacity=0.2,draw=blue] (8.4,1.9) -- (9.1,1.9) -- (9.1,2.6) -- (8.4,2.6) -- (8.4,1.9);

 		\node (d-u) at (8.9,1.7){};
 		\node (d-l) at (9.2,1.2){};
		
         \path (d-u) -- (d-l) node [black, font=\Huge, midway,sloped] {$\dots$};		
		
  		\draw[fill=black,opacity=0.2,draw=blue] (8.9,0.1) -- (9.6,0.1) -- (9.6,0.8) -- (8.9,0.8) -- (8.9,0.1);
  		\draw[fill=black,opacity=0.2,draw=blue] (9.0,0) -- (9.7,0) -- (9.7,0.7) -- (9,0.7) -- (9,0);

 		\draw (9.4,0.5) -- (9.5,0.5) -- (9.5,0.6) -- (9.4,0.6) -- (9.4,0.5);
		
  		\draw (9.45,0.6) -- (10.3,0.3);
  		\draw (9.45,0.5) -- (10.3,0.3);

 		\draw[fill=black,opacity=0.2,draw=blue] (9.3,2) -- (9.8,2) -- (9.8,2.5) -- (9.3,2.5) -- (9.3,2);
 		\draw[fill=black,opacity=0.2,draw=blue] (9.4,1.9) -- (9.9,1.9) -- (9.9,2.4) -- (9.4,2.4) -- (9.4,1.9);

  		\node (d-u) at (9.8,1.5){};
  		\node (d-l) at (10.1,1.0){};
		
          \path (d-u) -- (d-l) node [black, font=\Huge, midway,sloped] {$\dots$};		
		
  		\draw[fill=black,opacity=0.2,draw=blue] (9.9,0.1) -- (10.4,0.1) -- (10.4,0.6) -- (9.9,0.6) -- (9.9,0.1);
  		\draw[fill=black,opacity=0.2,draw=blue] (10,0) -- (10.5,0) -- (10.5,0.5) -- (10,0.5) -- (10,0);

  		\draw (7.3,0.5) -- (7.4,0.5) -- (7.4,0.6) -- (7.3,0.6) -- (7.3,0.5);
		
  		\draw (7.35,0.6) -- (8.3,0.3);
  		\draw (7.35,0.55) -- (8.3,0.3);

 	    \node (conv-left) at (7.,-0.5){};
 		\node (conv-right) at (9.3,-0.5){};
 		\draw[] (conv-left) -- (conv-right);
 		\node[align=center] at (8.1,-1){$10\times$ Residual \\Block };

 	    \node (conv-left) at (9.5,-0.5){};
 		\node (conv-right) at (11.,-0.5){};
 		\draw[] (conv-left) -- (conv-right);
 		\node[align=center] at (10.3,-1.25){Conv + \\ Global\\ Pooling };

		
  		\draw[fill=black,opacity=0.2,draw=blue] (10.3,2) -- (10.6,2) -- (10.6,2.3) -- (10.3,2.3) -- (10.3,2.0);
  		\draw[fill=black,opacity=0.2,draw=blue] (10.4,1.9) -- (10.7,1.9) -- (10.7,2.2) -- (10.4,2.2) -- (10.4,1.9);
  		\draw[fill=black,opacity=0.2,draw=blue] (10.5,1.8) -- (10.8,1.8) -- (10.8,2.1) -- (10.5,2.1) -- (10.5,1.8);
  		\draw[fill=black,opacity=0.2,draw=blue] (10.6,1.7) -- (10.9,1.7) -- (10.9,2.) -- (10.6,2.) -- (10.6,1.7);

  		\node (d-u) at (10.9,1.4){};
  		\node (d-l) at (11.3,0.9){};
		
          \path (d-u) -- (d-l) node [black, font=\Huge, midway,sloped] {$\dots$};

  		\draw[fill=black,opacity=0.2,draw=blue] (11.3,0.3) -- (11.6,0.3) -- (11.6,0.6) -- (11.3,0.6) -- (11.3,0.3);
  		\draw[fill=black,opacity=0.2,draw=blue] (11.4,0.2) -- (11.7,0.2) -- (11.7,0.5) -- (11.4,0.5) -- (11.4,0.2);

  		\draw[fill=black,opacity=0.2,draw=blue] (11.5,0.1) -- (11.8,0.1) -- (11.8,0.4) -- (11.5,0.4) -- (11.5,0.1);
  		\draw[fill=black,opacity=0.2,draw=blue] (11.6,0.0) -- (11.9,0.0) -- (11.9,0.3) -- (11.6,0.3) -- (11.6,0.0);

  	    \node (conv-left) at (11.1,-0.5){};
  		\node (conv-right) at (12.8,-0.5){};
  		\draw[] (conv-left) -- (conv-right);
  		\node[align=center] at (12.,-1.25){Fully \\Connected \\Layer};

   		\draw[fill=black,opacity=0.2,draw=blue] (13.5,1.7) -- (13.8,1.7) -- (13.8,2) -- (13.5,2) -- (13.5,1.7);		
   		\draw[fill=black,opacity=0.2,draw=blue] (13.5,1.3) -- (13.8,1.3) -- (13.8,1.6) -- (13.5,1.6) -- (13.5,1.3);		
		
   		\draw[fill=black,opacity=0.2,draw=blue] (13.5,0.9) -- (13.8,0.9) -- (13.8,1.2) -- (13.5,1.2) -- (13.5,0.9);		

   		\draw[fill=black,opacity=0.2,draw=blue] (13.5,0.5) -- (13.8,0.5) -- (13.8,0.8) -- (13.5,0.8) -- (13.5,0.5);		

    \node[inner sep=0pt] (im1e09) at (13.65,1.85)
    {\includegraphics[width=.4cm]{Architectures/im_1e9.png}};

     \node[inner sep=0pt] (im1e08) at (13.65,1.45)
    {\includegraphics[width=.4cm]{Architectures/im_1e8.png}};
   
    \node[inner sep=0pt] (im1e07) at (13.65,1.05)
    {\includegraphics[width=.4cm]{Architectures/im_1e7.png}};

     \node[inner sep=0pt] (im1e06) at (13.65,0.65)
     {\includegraphics[width=.4cm]{Architectures/im_1e6.png}};

 	    \node (conv-left) at (13.,-0.5){};
 		\node (conv-right) at (14.5,-0.5){};
 		\draw[] (conv-left) -- (conv-right);
 		\node[align=center] at (13.8,-1){Output};

	\end{tikzpicture}